\documentclass[useAMS,usenatbib]{mn2e}
\usepackage{graphicx}
\usepackage{amssymb}

\def\msun{M_\odot}
\def\fs{f_{\rm S}}
\def\fsi{f_{\rm S_I}}

\def\I0{I_{\rm 0}}
\def\V0{V_{\rm 0}}

\def\tE{t_{\rm E}}
\def\te{t_{\rm E}}
\def\t0{t_{\rm 0}}
\def\u0{u_{\rm 0}}

\newcommand{\eg}{{e.g.},\,}
\newcommand{\ie}{{i.e.},\,}
\newcommand{\apj}{{Astrophysical Journal}}
\newcommand{\apjl}{{Astrophysical Journal Letters}}
\newcommand{\apjs}{{Astrophysical Journal Supplement Series}}
\newcommand{\araa}{{ARA\&A}}
\newcommand{\aap}{{Astronomy \& Astrophysics}}
\newcommand{\mnras}{{MNRAS}}

\title[The OGLE-III View of Microlensing towards the SMC]
{The OGLE View of Microlensing towards the Magellanic Clouds. IV. OGLE-III SMC Data and Final Conclusions on MACHOs.\thanks{Based on
 observations obtained with the 1.3~m Warsaw telescope at the Las Campanas Observatory of the Carnegie Institution of Washington.}}
\author[{\L}. Wyrzykowski et al.]
{{\L}. Wyrzykowski$^{1,2}$\thanks{email: wyrzykow@ast.cam.ac.uk, name
 pronunciation: {\it Woocash Vizhikovsky}}, J. Skowron$^{2,3}$, S. Koz{\l}owski$^{2,3}$,
A. Udalski$^2$, M. K. Szyma{\'n}ski$^2$, \newauthor
M. Kubiak$^2$,  G. Pietrzy{\'n}ski$^{2,4}$, I. Soszy{\'n}ski$^2$, O. Szewczyk$^{2,4}$, \newauthor
K. Ulaczyk$^2$, R. Poleski$^2$, P. Tisserand$^5$\\
$^1$ Institute of Astronomy, University of Cambridge,  Madingley~Road,  Cambridge~CB3~0HA,~UK \\
$^2$ Warsaw University Astronomical Observatory, Al.~Ujazdowskie~4, 00-478~Warszawa, Poland \\
$^3$ Department of Astronomy, The Ohio State University, 140 W. 18th Ave., Columbus, OH 43210, USA\\
$^4$ Universidad de Concepci{\'o}n, Departamento de Astronomia, Casilla 160-C, Concepci{\'o}n, Chile\\
$^5$ Research School of Astronomy and Astrophysics, Australian National University, Cotter Rd, Weston Creek ACT 2611, Australia\\
}

\begin{document}

\date{Accepted 2011 June 13. Received 2011 June 13; in original form 2011 March 23}

\pagerange{\pageref{firstpage}--\pageref{lastpage}} \pubyear{2011}

\maketitle

\label{firstpage}

\begin{abstract}

In this fourth part of the series presenting the Optical Gravitational Lensing Experiment (OGLE) microlensing studies of the dark matter halo compact objects (MACHOs) we describe results of the OGLE-III monitoring of the Small Magellanic Cloud (SMC). Three sound candidates for microlensing events were found and yielded the optical depth $\tau_{\rm SMC-OIII}=1.30\pm1.01~\times~10^{-7}$, 
 consistent with the expected contribution from Galactic disk and SMC self-lensing.
We report that event OGLE-SMC-03 is the most likely a thick disk lens candidate, the first of such type found towards the SMC.
In this paper we also combined all OGLE Large and Small Magellanic Cloud microlensing results in order to refine the conclusions on MACHOs.
All but one of OGLE events are most likely caused by the lensing by known populations of stars, therefore we concluded that there is no need for introducing any special dark matter compact objects in order to explain the observed events rates.
Potential black hole event indicates that similar lenses can contribute only about 2 per cent to the total mass of the halo, which is still in agreement with the expected number of such objects.

\end{abstract}

\begin{keywords}
Cosmology: Dark Matter, Gravitational Lensing, Galaxy: Structure, Halo, Galaxies: Small Magellanic Cloud
\end{keywords}

\section{Introduction}
Campaigns like OGLE, MACHO or EROS were all founded nearly 20 years ago primarily to settle the question whether dark matter resided in the galactic halo in the form of compact objects (MACHOs). 
The background theory of the method employing gravitational microlensing was provided by Bohdan Paczy{\'n}ski \citep{Paczynski1986} and in subsequent years Paczy{\'n}ski himself advocated and supported the creation of large-scale photometric surveys towards the Magellanic Clouds.
The Large and Small Magellanic Clouds (LMC and SMC) were considered the best targets for such studies, as with a moderate telescope size the stars of the Clouds were easily resolved and hence formed a perfect background for the hunt for microlensing events.
These are unique and temporal brightening events observed on a source background star caused by the gravitational lensing phenomenon in which the lensing object is located in front of the background star.

In the most generic case of a point lens mass and a point source the gravitational amplification of the background light is described by equation (after \citealt{Paczynski1996}):
\begin{equation}
\label{eq:A}
A= { u^2 + 2 \over u\sqrt{u^2+4} } \qquad {\rm and} \qquad u= \sqrt{\u0^2 + {{(t-\t0)^2} \over {\tE^2}}},
\end{equation}

\noindent
where $\t0$ is the time of the maximum of the peak, $\tE$ is the Einstein radius crossing time (event time-scale) and $\u0$ is the event impact parameter.

Parameter $\tE$ is the only parameter linked with the physical properties of the source-lens system and is dependent on the lens mass, the lens and source distances and the relative velocity between lens and source.
This means that, in principle, a single microlensing event can not provide any information regarding the lensing object. Additional data are possible only in case of non-standard events, like binary lens/source or those showing a parallax effect.
Hence, in studying the mass distribution of lenses, microlensing events are regarded only in groups.
The statistical value related to the total mass enclosed in the volume towards the background sources is the microlensing optical depth, $\tau$:
\begin{equation}
\label{eq:tau}
\tau={{\pi}\over {2 N_* T_{\rm obs}}}\displaystyle \sum_i^{N_{\rm ev}} {{\tE}_{i} \over \epsilon({\tE}_i)}
\end{equation}
where $T_{\rm obs}$ is the time-span of all observations,
$N_*$ is the total number of monitored stars,
$N_{\rm ev}$ is the total number of events, each with time-scale ${\tE}_{i}$ and detection efficiency $\epsilon({\tE}_i)$.

In the case of the Magellanic Clouds the overall optical depth can be composed with contributions from lenses at various levels. 
Apart from the hypothetical dark matter compact objects there can also be a contribution to the optical depth from luminous lenses from the disk of our Galaxy and from the frontal side of the Cloud itself (\ie self-lensing, SL).

Observations and studies of the pixel-lensing microlensing events towards M31 are still not conclusive with respect to the amount of self-lensing and MACHO lenses (see \citealt{CalchiNovati2010review} for recent review). 
However, the most recent analysis of the data tend to be inclined towards self-lensing as being the only explanation of the observed signal \citep{CalchiNovati2010}.

The most recent studies of the data gathered by microlensing surveys towards the LMC suggest the contribution from dark matter compact objects with masses below 10 $\msun$ can be practically ruled out. 
The upper limit on MACHO abundance in the halo was set to 6-7 per cent by the EROS \citep{TisserandEROSLMC} and OGLE groups \citep{OGLE3LMC} (see \citealt{Moniez2010} for review).
The core of the detected microlensing signal seems to come from the self-lensing events, as suggested already in \citet{Sahu1994a} and \citet{Sahu1994b}.

The LMC self-lensing effect is relatively well understood thanks to the convenient, almost face-on alignment of the LMC towards us. 
The studies by \cite{Mancini2004}, \cite{CalchiNovati2009} and \cite{CalchiNovati2011} showed that the average self-lensing optical depth is in the order of $0.4\times10^{-7}$, which is consistent with what was measured in OGLE-II (\citealt{OGLE2LMC}, hereafter Paper I) and OGLE-III data (\citealt{OGLE3LMC}, hereafter Paper III). 

The situation with the SMC is more complicated. This dwarf galaxy seems to be elongated and is stretched out along the line-of-sight, which produces a much higher self-lensing optical depth. 
Some estimates by \cite{EROSSMC1998} and \cite{Graff1999}
place this value at at least $1.0 \times 10^{-7}$, but it can actually be much higher.
Microlensing studies of the SMC tend to confirm this with their typically higher event rates (\citealt{TisserandEROSLMC}, \citealt{OGLE2SMC}, hereafter Paper II), compared with the LMC.

In this paper we conclude the series of studies of the OGLE microlensing data towards the both Magellanic Clouds with the analysis of the 8 years of observations of the SMC in the course of the OGLE-III project.
We attempt to finalise the investigation of the subtle topic of dark matter compact halo objects, providing the result from currently the best and the most suitable data set.

The paper is organised as follows. First, the OGLE-III data used for the analysis are presented. Then, we describe the method applied for finding the microlensing events and its results with a detailed description and study of each found event. This is followed by a calculation of the optical depth and discussion of the results.

\section{Observational Data}
\label{sec:data}

The photometric data used in this study were collected during the third phase of the OGLE project (2001-2009)
with the 1.3-m Warsaw telescope located at Las Campanas Observatory,
Chile, operated by the Carnegie Institution of Washington.
The ``second generation'' camera comprised eight SITe
$2048\times4096$ CCD detectors with 15~$\mu$m pixels resulted in
0.26~arcsec/pixel and $35\times35$~arcmins total field of view.
The details on the instrumentation setup can be found in \citep{Udalski2003}.

\begin{figure*}
\includegraphics[width=13cm]{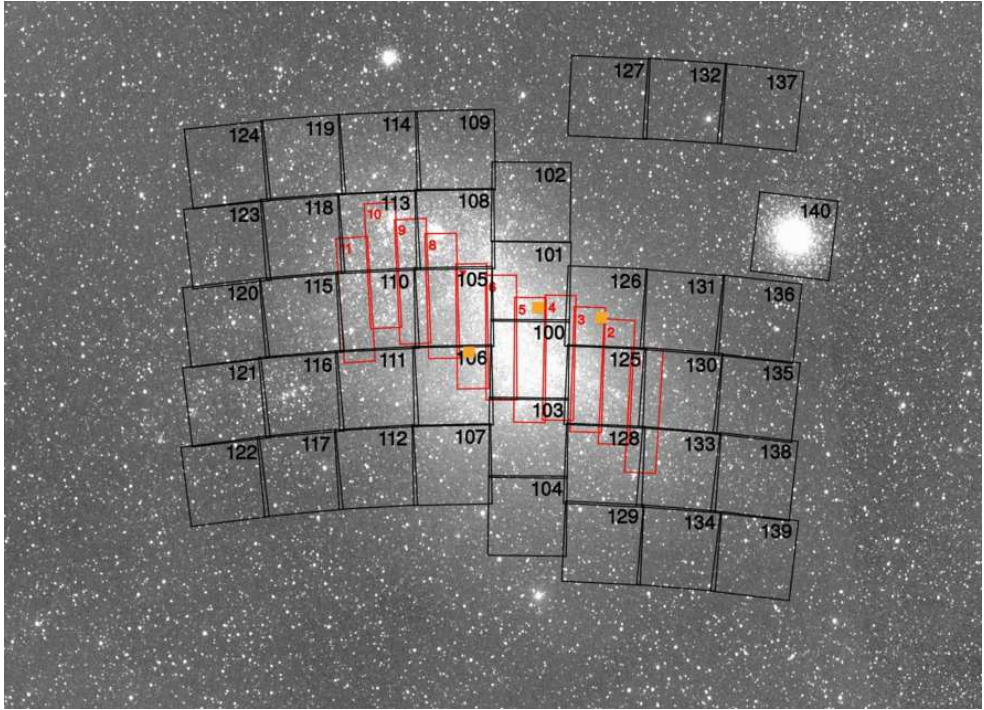}
\caption{Positions of the OGLE--III SMC fields (black). Also shown are all OGLE--II fields (red rectangles).
The three small filled squares show  the positions of the HST fields used for our blending determination.
Background image credit: ASAS all sky survey.}
\label{fig:fields}
\end{figure*}

The central regions and the outskirts of the SMC were covered by 41 fields, giving a total of 14 square degrees.
The map of the fields is shown in Fig. \ref{fig:fields}.
The statistical details of the fields are gathered in Table \ref{tab:fields} with the following information: the coordinates of their centers,
the number of ``good'' template objects in the $I$-band, the blending-corrected number of stars
(see Section \ref{sec:blending}) and the mean number of all objects visible on a single CCD (of 8) used for assigning the blending density level.
``Good'' objects are the template objects with at least 80 good data points
(excluding measurements with very large error-bars) and mean magnitude brighter than $21.0$ mag (chosen as a mean peak of the observed luminosity functions).  

The very first observations of the SMC within the OGLE--III phase were taken in June 2001 (JD=$2\;452\;085$), except field SMC140 (started in July 2004), and continued until May 2009 (JD=$2\;454\;954$).
Most of the observations were taken through the Cousins $I$-band filter with exposure time 180~s with a mean seeing of 1.36 arcsec.
Between 583 and 762 measurements were gathered in each field with an average sampling varying from 2.0 to 3.0 days between subsequent frames (excluding the gaps between the seasons). The only exception is field SMC128 which was observed 1228 times with an average sampling of 1.6 days.
Additionally, between 47 and 114 observations per field were obtained in Johnson $V$-band with integration time 225~s and a mean seeing of 1.39 arcsec.
First $V$-band observations were taken in November 2004 (JD=$2\;453\;314$) and since then $V$ filter frames were taken regularly.
The average sampling frequency in the $V$-band was between 4.6 and 10.3 days.

The image reduction pipeline used an image subtraction technique and was based on Difference Image Analysis (DIA; \citealt{AlardLuptonDIA}, \citealt{WozniakDIA}).
The data used in this work come from the final reductions calibrated to the standard system.
The template images used for subtraction were composed from selected best individual frames resulting in a seeing of around 0.9 arcsec and 1.0 arcsec for I-band and V-band, respectively.
The full description of the reduction techniques, photometric calibration and astrometric transformations can be found in \citep{Udalski2008OGLE3}.

Photometric errors produced by the difference imaging process were adjusted as described in Paper I by deriving a magnitude dependent correction factor using the constant stars. 
For each SMC field and each filter ($f=I,~V$) we found parameters $\gamma_f$ and $\epsilon_f$, which correct the original error-bar according to the formula:
\begin{equation}
\sigma_{\rm mag_f~cor} = \sqrt{ ( \gamma_f \sigma_{\rm mag_f})^2 + \epsilon_f^2 },
\label{eq:errors}
\end{equation}
where $\sigma_{\rm mag}$ is the original error-bar returned by the DIA.

Error-correction parameters averaged over all SMC fields yielded with 
$\langle\gamma_I\rangle = 1.04031$,
$\langle\epsilon_I\rangle = 0.00414852$,
$\langle\gamma_V\rangle= 0.867849$,
$\langle\epsilon_V\rangle = 0.0026929$. 
Table \ref{tab:errorcor} shows the error correction coefficients for the first few fields. The full table can be found on the OGLE website\footnote{http://ogle.astrouw.edu.pl/}.

\begin{table*}
\centering
\caption{OGLE--III SMC fields.}
\label{tab:fields}
\begin{tiny}
\begin{tabular}{cccrrc|cccrrc}
\hline
\noalign{\vskip5pt}
Field & $\alpha_{J2000}$ & $\delta_{J2000}$ & \multicolumn{2}{c}{$N_{*\mathrm{good}}$[$10^3$]} &
$\langle \log{N_{\mathrm{all_{CCD}}}}\rangle$ &
Field & $\alpha_{J2000}$ & $\delta_{J2000}$ & \multicolumn{2}{c}{$N_{*\mathrm{good}}$[$10^3$]} &
$\langle \log{N_{\mathrm{all_{CCD}}}}\rangle$ \\
& & & tpl & real &  & & & & tpl & real &  \\
SMC100 & 0:50:02.9 & -73:08:40 & 469.8 & 550.5 & 4.81 & SMC121 & 1:22:01.2 & -73:20:35 & 28.7 & 29.2 & 3.60 \\
SMC101 & 0:50:00.4 & -72:33:18 & 315.5 & 353.6 & 4.63 & SMC122 & 1:23:12.8 & -73:56:05 & 19.3 & 19.6 & 3.43 \\
SMC102 & 0:50:05.0 & -71:57:26 & 106.4 & 111.3 & 4.18 & SMC123 & 1:20:02.3 & -72:09:34 & 31.3 & 31.9 & 3.64 \\
SMC103 & 0:50:04.8 & -73:44:00 & 249.9 & 274.8 & 4.54 & SMC124 & 1:19:09.6 & -71:33:44 & 41.0 & 41.9 & 3.74 \\
SMC104 & 0:50:05.0 & -74:19:23 & 89.9 & 94.0 & 4.21 & SMC125 & 0:42:04.2 & -73:19:45 & 309.8 & 351.3 & 4.68 \\
SMC105 & 0:57:49.6 & -72:44:41 & 338.0 & 378.3 & 4.66 & SMC126 & 0:42:21.8 & -72:44:08 & 189.2 & 203.2 & 4.41 \\
SMC106 & 0:58:05.5 & -73:20:17 & 314.6 & 352.7 & 4.66 & SMC127 & 0:42:54.6 & -71:09:03 & 19.2 & 19.5 & 3.43 \\
SMC107 & 0:58:21.9 & -73:55:54 & 125.0 & 131.3 & 4.25 & SMC128 & 0:41:46.2 & -73:55:25 & 172.0 & 183.4 & 4.38 \\
SMC108 & 0:57:32.5 & -72:09:24 & 379.9 & 434.5 & 4.71 & SMC129 & 0:41:31.9 & -74:30:42 & 50.4 & 51.7 & 3.85 \\
SMC109 & 0:57:20.9 & -71:33:48 & 111.1 & 116.6 & 4.22 & SMC130 & 0:34:06.1 & -73:19:39 & 114.0 & 119.0 & 4.20 \\
SMC110 & 1:05:35.8 & -72:44:45 & 198.1 & 213.0 & 4.47 & SMC131 & 0:34:34.0 & -72:44:11 & 65.9 & 68.0 & 3.95 \\
SMC111 & 1:06:06.0 & -73:20:14 & 161.8 & 171.9 & 4.36 & SMC132 & 0:35:49.7 & -71:08:46 & 16.7 & 17.0 & 3.37 \\
SMC112 & 1:06:40.9 & -73:55:48 & 92.3 & 96.1 & 4.15 & SMC133 & 0:33:36.9 & -73:55:24 & 94.2 & 97.8 & 4.10 \\
SMC113 & 1:05:03.2 & -72:09:28 & 241.6 & 264.7 & 4.55 & SMC134 & 0:32:52.5 & -74:30:42 & 39.4 & 40.2 & 3.73 \\
SMC114 & 1:04:36.8 & -71:33:42 & 145.4 & 153.5 & 4.29 & SMC135 & 0:26:03.8 & -73:19:45 & 46.4 & 47.6 & 3.83 \\
SMC115 & 1:13:19.4 & -72:44:42 & 87.9 & 91.4 & 4.14 & SMC136 & 0:26:48.7 & -72:44:14 & 36.3 & 37.1 & 3.71 \\
SMC116 & 1:14:05.2 & -73:20:08 & 107.4 & 112.0 & 4.17 & SMC137 & 0:28:42.4 & -71:08:44 & 14.1 & 14.3 & 3.30 \\
SMC117 & 1:15:00.4 & -73:55:44 & 37.2 & 38.1 & 3.77 & SMC138 & 0:25:21.4 & -73:55:17 & 53.8 & 55.4 & 3.99 \\
SMC118 & 1:12:35.3 & -72:09:12 & 72.6 & 75.2 & 4.06 & SMC139 & 0:24:16.5 & -74:30:48 & 30.5 & 31.1 & 3.63 \\
SMC119 & 1:11:54.1 & -71:33:40 & 96.6 & 100.1 & 4.11 & SMC140 & 0:24:05.1 & -72:05:00 & 301.6 & 341.8 & 4.63 \\
SMC120 & 1:21:05.2 & -72:44:46 & 55.5 & 56.9 & 3.87 &  &  &  &  &  &  \\
\noalign{\vskip5pt}
\hline
\noalign{\vskip5pt}
\hline
\noalign{\vskip5pt}
\multicolumn{9}{r}{total}      & 5470352 & 5971776 & \\
\noalign{\vskip5pt}
\hline
\end{tabular}
\end{tiny}
\medskip
\begin{flushleft}
{\it Note:} Coordinates point to the centre of the field (centre of the mosaic), each being $35' \times
35'$. Number of ``good'' objects in the template is provided ($N>80$ and
$\langle I \rangle < 21.0$ mag) together with the estimated number of
real monitored stars (see Section \ref{sec:blending}).
Mean number of all objects detected on a single CCD used for calculating the density of a field is given in the last column.
\end{flushleft}
\medskip
\bigskip
\bigskip
\end{table*}

\begin{table}
\centering
\caption{Error correction coefficients for each CCD chip of the first four OGLE--III LMC fields for $I$- and $V-$ bands. The full table is available on-line from the OGLE website.}
\label{tab:errorcor}
\begin{tiny}
\begin{tabular}{ccccc}
\hline
Field & $\gamma_\mathrm{I}$ & $\epsilon_\mathrm{I}$ & $\gamma_\mathrm{V}$ & $\epsilon_\mathrm{V}$ \\
\hline
SMC100.1 & 0.892 & 0.0040 & 0.787 & 0.0029 \\
SMC100.2 & 0.903 & 0.0044 & 0.764 & 0.0026 \\
SMC100.3 & 0.891 & 0.0042 & 0.736 & 0.0027 \\
SMC100.4 & 0.930 & 0.0037 & 0.793 & 0.0026 \\
SMC100.5 & 0.911 & 0.0048 & 0.758 & 0.0028 \\
SMC100.6 & 0.922 & 0.0043 & 0.780 & 0.0026 \\
SMC100.7 & 0.921 & 0.0044 & 0.787 & 0.0028 \\
SMC100.8 & 1.026 & 0.0044 & 0.840 & 0.0032 \\
SMC101.1 & 0.947 & 0.0040 & 0.963 & 0.0029 \\
SMC101.2 & 0.939 & 0.0044 & 0.895 & 0.0026 \\
SMC101.3 & 0.951 & 0.0042 & 0.916 & 0.0027 \\
SMC101.4 & 1.000 & 0.0037 & 1.138 & 0.0026 \\
SMC101.5 & 1.010 & 0.0048 & 1.057 & 0.0028 \\
SMC101.6 & 1.043 & 0.0043 & 1.100 & 0.0026 \\
SMC101.7 & 1.002 & 0.0044 & 1.050 & 0.0028 \\
SMC101.8 & 1.145 & 0.0044 & 1.089 & 0.0032 \\
SMC102.1 & 0.944 & 0.0040 & 0.827 & 0.0029 \\
SMC102.2 & 0.951 & 0.0044 & 0.816 & 0.0026 \\
SMC102.3 & 0.972 & 0.0042 & 0.830 & 0.0027 \\
SMC102.4 & 0.977 & 0.0037 & 0.901 & 0.0026 \\
SMC102.5 & 0.951 & 0.0048 & 0.839 & 0.0028 \\
SMC102.6 & 1.030 & 0.0043 & 0.901 & 0.0026 \\
SMC102.7 & 1.005 & 0.0044 & 0.908 & 0.0028 \\
SMC102.8 & 1.109 & 0.0044 & 0.869 & 0.0032 \\
SMC103.1 & 0.995 & 0.0040 & 0.925 & 0.0029 \\
SMC103.2 & 1.002 & 0.0044 & 1.015 & 0.0026 \\
SMC103.3 & 0.996 & 0.0042 & 0.973 & 0.0027 \\
SMC103.4 & 1.029 & 0.0037 & 0.955 & 0.0026 \\
SMC103.5 & 1.019 & 0.0048 & 0.885 & 0.0028 \\
SMC103.6 & 1.024 & 0.0043 & 0.977 & 0.0026 \\
SMC103.7 & 1.042 & 0.0044 & 1.015 & 0.0028 \\
SMC103.8 & 1.121 & 0.0044 & 0.962 & 0.0032 \\
\dots & & & &  \\
\hline
\end{tabular}
\end{tiny}
\end{table}

\section{Search procedure}
\label{sec:search}

For the SMC OGLE-III data we applied the same automated pipeline for detecting microlensing events as designed and used in searching through the LMC OGLE-II (Paper I), SMC OGLE-II (Paper II) and LMC OGLE-III (paper III) data. The procedure for the SMC contains 10 cuts applied to the pre-selected ``good'' data (with enough good observations) and limited to the right magnitude level. 
Table \ref{tab:conditions} describes all cuts applied to the data for two star samples (defined at cut 0): All Stars - limited to 21 mag (maximum of the observed luminosity functions, \ie the end of OGLE completeness in these fields) and Bright Stars - limited to 19.3 mag (1 mag below mean Red Clump position).

In the first cut for each star sample we selected light curves with a significant bump over baseline, following the concept of \citet{SumiOGLEbulge}. Then, we removed all candidates with their baseline colour and magnitude located in the ``blue bumper'' region to exclude these most common contaminants related with the Be stars. 
In cut 3 we chose those light curves for which the standard microlensing fit (with blending parameter fixed, hence index $\mu4$ in the table \ref{tab:conditions}) was better than a constant line fit. 
In the following cut we required that there were at least 6 data points in the microlensing peak between $\t0 - \tE$ and $\t0 + \tE$. 

In cut 5 we fitted the candidate bump also with the simple supernova model (exponential decline) and selected only those for which the $\chi^2$ of the microlensing fit was better (either with blending or without). The number of expected supernovae in the entire OGLE-III coverage of the SMC is around 22 (after \citealt{Dilday2010}, assuming 20 per cent detection efficiency), and indeed many of the objects removed at this stage resembled supernovae outbursts, some clearly with the host galaxy visible on the OGLE image. 
In this way we also removed all other asymmetric bumps.
The objects remaining after this step were visually inspected and most of them turned out to be either nova-like events or red-baseline bumpers.

We then checked if the peak of the candidate bump was located within the data span (cut 6).
In the next two cuts (7 and 8) we required that the blended microlensing model fit converged and yielded reasonable values, allowing for some negative blending (effect of instrumental origin, see \eg \citealt{Smith2007blending}), and that the $\chi^2/N_\mathrm{dof}$ of the blended fit was better than 2.6 and for non-blended fit around the peak better than 4.5.

Finally, we restricted our search of microlensing events to the ones with time-scales between 1 and 1000 days and with impact parameters below 1. 

\begin{table*}
\centering
\caption{Selection criteria for the search for microlensing events in the OGLE--III SMC data and the number of objects left after each cut for the All Stars and Bright Stars Samples.}
\label{tab:conditions}
\begin{tabular}[h]{c|lrrr}
\hline
Cut no. & & & \multicolumn{2}{c}{No. of objects left} \\
         & & & All  &  Bright \\
\hline     
0a & Selection of ``good'' objects & $N>80$, $\langle I \rangle \le 21.0$ mag  &  5,470,352 & \\
0b &                                          & $N>80$, $\langle I \rangle \le 19.3$ mag &                  & 1,614,327 \\
& & & &\\

1 & Significant bump over baseline & $\displaystyle\sum_\mathrm{peak} \sigma_i > 30.0 $ & 1,217 & 1,060\\
& & & & \\

2 & ``Bumper'' cut$^\dagger$ & $\langle I \rangle>19.0$~mag,  $\langle V-I \rangle>0.5$~mag & 482 & 380\\
& & & & \\

3 & Microlensing fit better than constant line fit & ${{\chi^2_\mathrm{line}-\chi^2_{{\mu} 4}}\over{{\chi^2_{{\mu}4}\over{N_{\mathrm{dof},\mu4}}}\sqrt{2N_{\mathrm{dof},\mu4,\mathrm{peak}}}}} > 100$  & 186 & 107\\
& & & & \\

4 & Number of points at the peak$^{\ast}$ & $N_\mathrm{peak} > 5$  & 182 & 104\\
& & & & \\

5 & Microlensing fit better than supernova fit, & $\chi^2_{SN} > \mathrm{MIN}(\chi^2,\chi^2_{{\mu}4})$ & 115 & 68 \\
& & & & \\

6 & Peak within the data span   & $2085 < {\t0} \le 4954$   & 112 & 65 \\
& [HJD-2450000]    &         &         \\
& & & & \\

7 & Blended fit converged & $0< \fs < 1.4$ & 35 & 23 \\
& & & & \\

8 & Conditions on goodness of microlensing fit &  ${{\chi^2}\over{N_\mathrm{dof}}} \le 2.6 $ and ${{\chi_{\mu4,\mathrm{peak}}^2}\over{N_{\mathrm{dof},\mu4,\mathrm{peak}}}} \le 4.5 $ & 7 & 6\\
& (global and at the peak) & & \\
& & & & \\

9 & Time-scale cut & $1 \le {\tE} \le 1000$ & 6 & 5 \\
& $[d]$ &  & \\
& & & & \\

10 & Impact parameter cut &  $0 < {\u0} \le 1 $ & 3 & 2 \\
& & & & \\

\hline
\end{tabular}
\\
\begin{flushleft}
$^{\dagger}$ magnitudes as in the field SMC100.1 (shifted according to the position of the center of Red Clump) \\
$^{\ast}$in the range of $\t0 \pm 1 \tE$ \\
\end{flushleft}
\end{table*}

\begin{figure*}
\includegraphics[width=10.0cm]{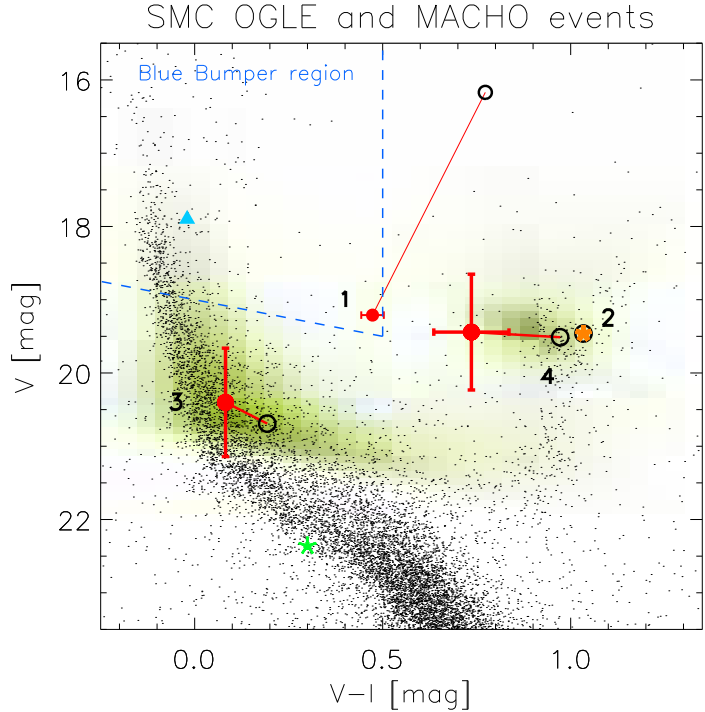}
\caption{Colour-magnitude diagram showing HST (black dots) and OGLE (green) stars of one of the OGLE-III fields. Red dots show estimated colour and brightness of the source in all OGLE events, whereas black open circles indicate the baseline of each event. Blue and green dots mark MACHO events from 1997 and 1998.
}
\label{fig:cmd}
\end{figure*}

\begin{figure*}
\includegraphics[width=10.0cm]{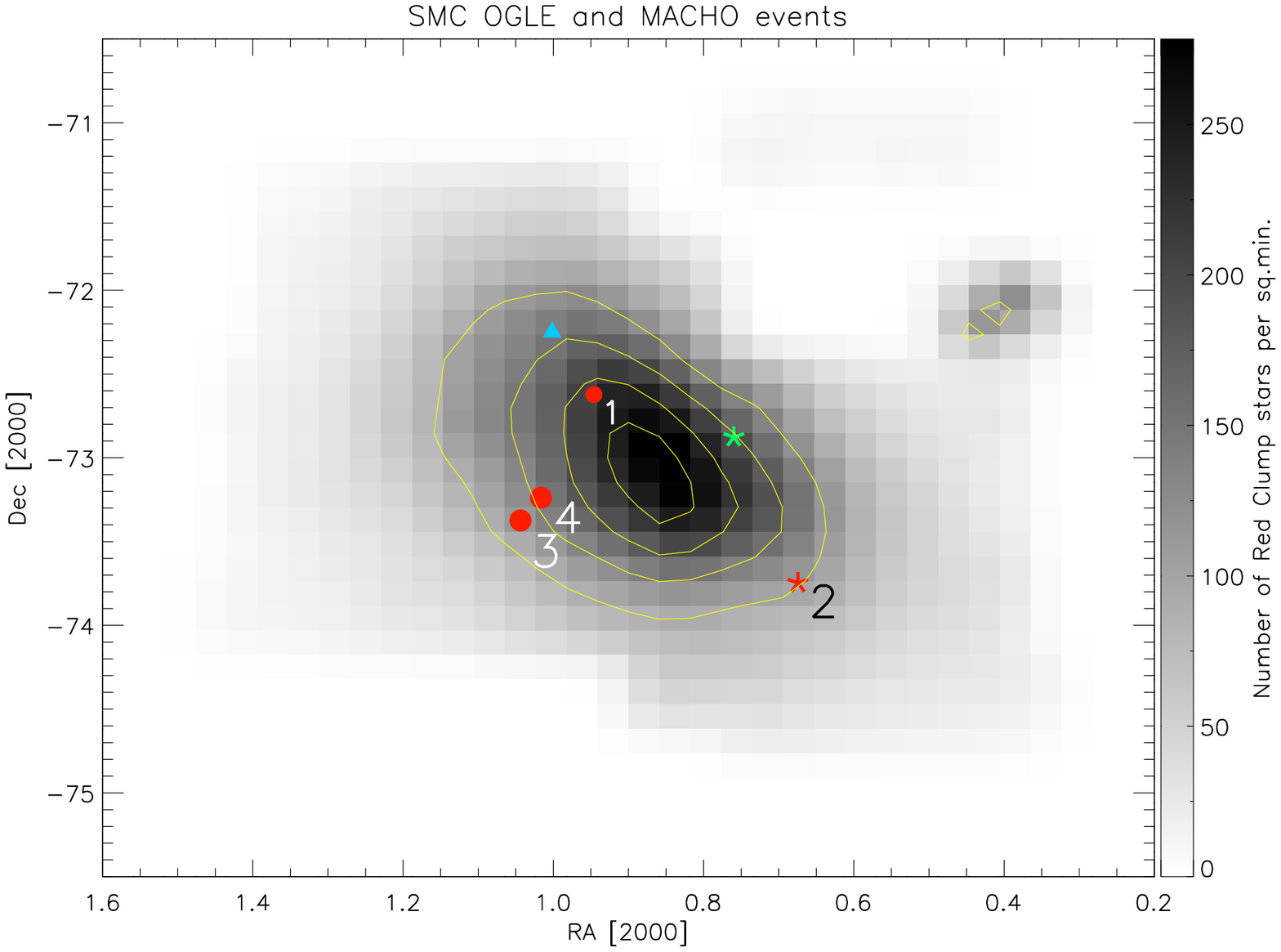}
\caption{Localisation of all SMC OGLE and MACHO events on the Red Clump stars density map. 
The colour coding is as that used in Figure \ref{fig:cmd}. Contours follow the density of Red Clump stars.  
}
\label{fig:mapevents}
\end{figure*}

\section{Results}
\label{sec:results}
Running our search pipeline on the SMC OGLE-III data found 3 candidates for microlensing events in All Stars Sample. 
Two of them were found in the Bright Stars Sample. 
The three events were dubbed OGLE-SMC-02, OGLE-SMC-03 and OGLE-SMC-04, with the numbers following the numbering started with the single candidate detected in the OGLE-II SMC data (Paper II). 

The basic information about these candidates are shown in Table \ref{tab:events}. 
First two events were detected while on-going by the Early Warning System (EWS) of OGLE, while the third one was not known before and occurred at the beginning of the OGLE-III when EWS was not yet operating.

\begin{figure*}
\center
\includegraphics[width=8.5cm]{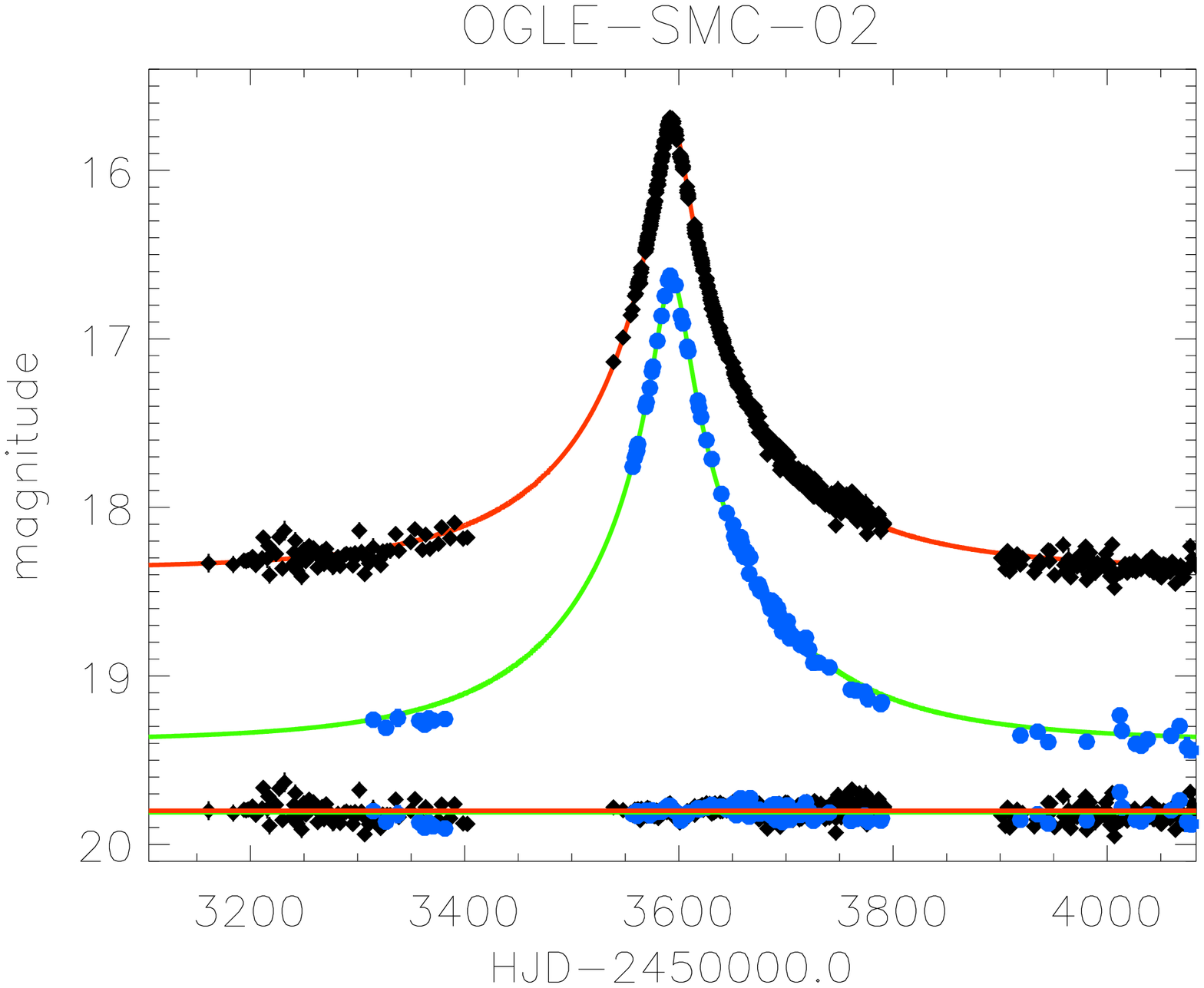}\\
\includegraphics[width=8.5cm]{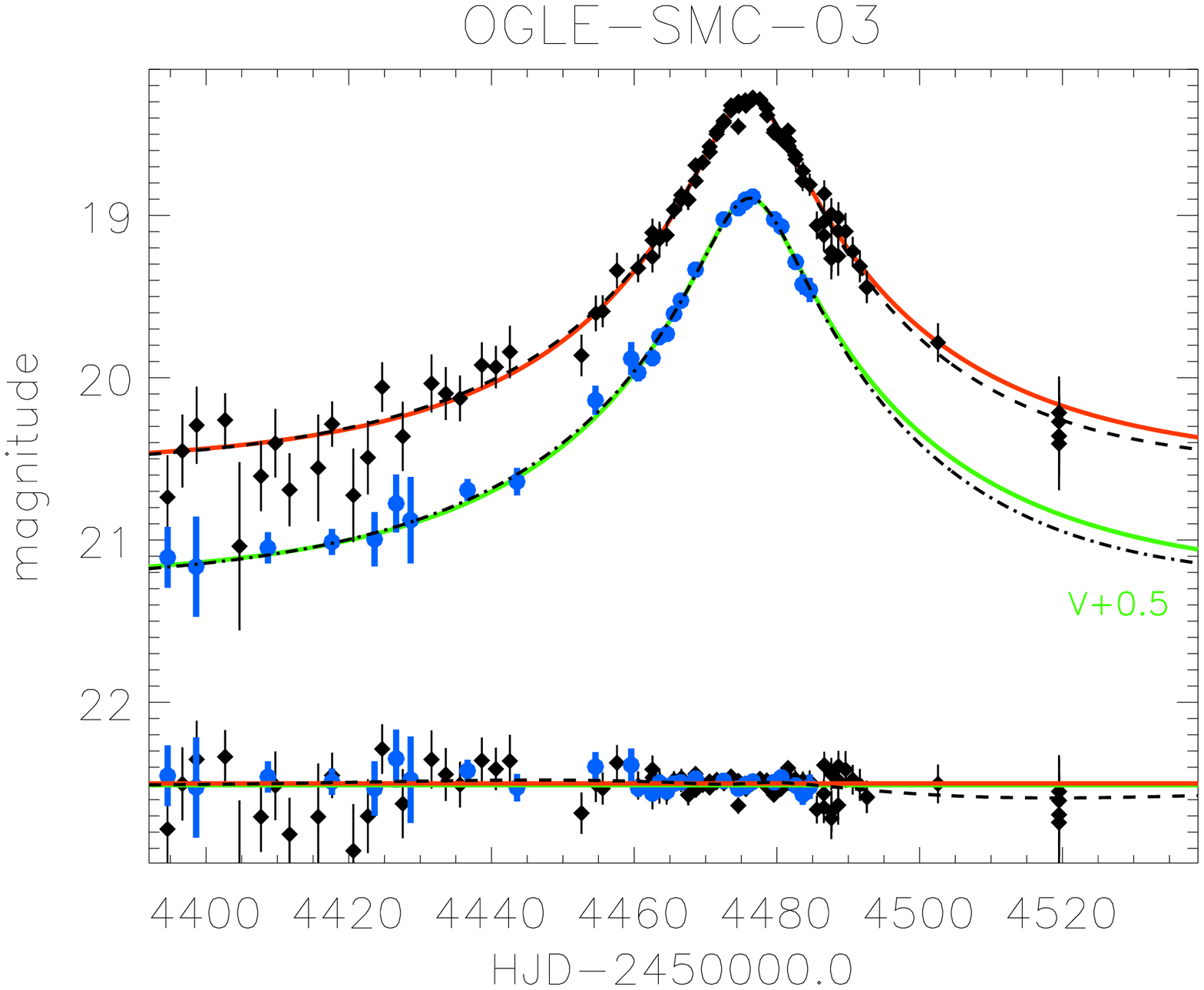}
\includegraphics[width=8.5cm]{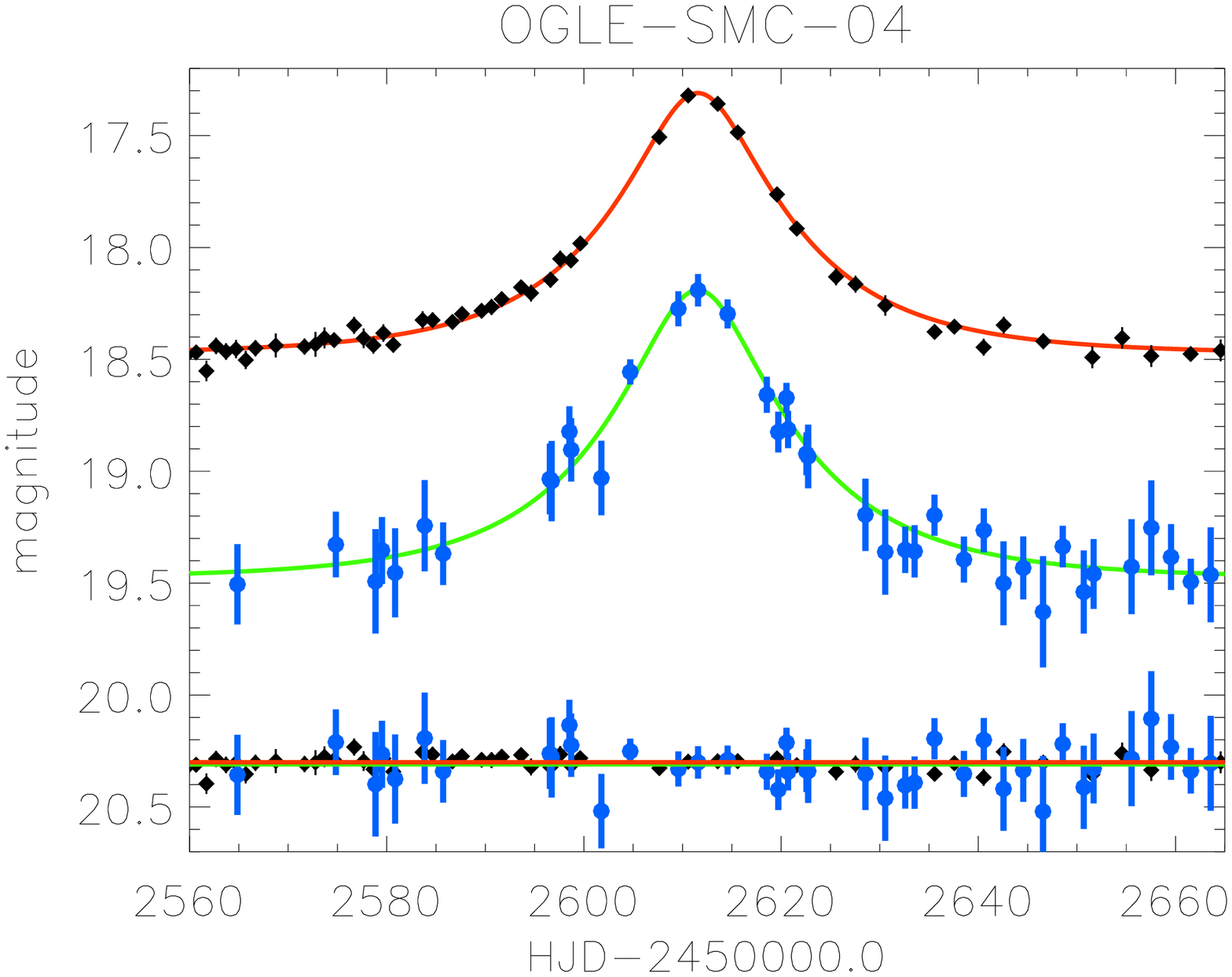}
\caption{Light curves and microlensing models of candidates for microlensing events detected in the OGLE--III SMC data. The standard microlensing model best-fitting to the data is shown in solid lines with black points and red curve for $I$-band, and blue points and green curve for $V$-band data (except OGLE-SMC-04, where it shows EROS $B$-band data). Dashed line shows parallax model fit. The residuals of the model fitting are shown as a respective lines and data points at the bottom of each panel.}
\label{fig:events}
\end{figure*}

\begin{figure*}
\center
\includegraphics[width=3.5cm]{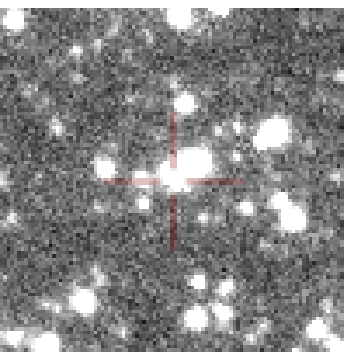}
\includegraphics[width=3.5cm]{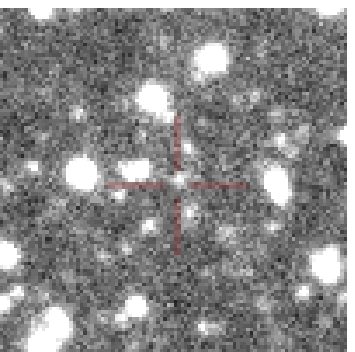}
\includegraphics[width=3.5cm]{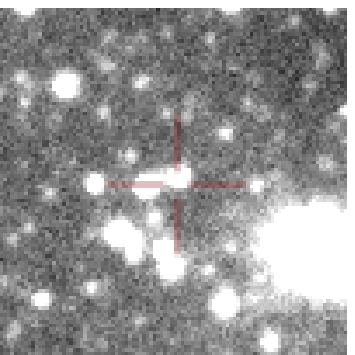}
\caption{Finding charts of the three candidates for microlensing events from OGLE-III SMC data. East is to the right, North is down. The side of each chart is 26 arc seconds.
Charts with events 02, 03 and 04 are shown from left to right.
A cross marks the object on which the microlensing brightening was detected.}
\label{fig:charts}
\end{figure*}

Figure \ref{fig:cmd} shows the colour-magnitude diagram with candidate events found in OGLE-II, OGLE-III and MACHO data. Figure \ref{fig:mapevents} presents a density of Red Clump stars of the SMC with marked positions of all candidate events. 
The light curves of three OGLE-III events with microlensing models fits are shown in Figure \ref{fig:events}, whereas their finding charts are presented in Figure \ref{fig:charts}.

Below we discuss each candidate in detail.

\begin{table*}
\caption{Microlensing events candidates detected in the OGLE--III SMC data. Events detected by the OGLE's Early Warning System (EWS) have their EWS designation given in the name column.}
\label{tab:events}
\begin{center}
\begin{tabular}{ccccccccc}
\hline
Event's name             & RA          & Dec       & field & db      &  baseline $I$      & baseline $V$ & source $I$ & source $V-I$\\
{\it (EWS)}                   & [J2000.0]   & [J2000.0] &         & star id & [mag]                   & [mag]                          & [mag] &   [mag]\\
& & & & & & & & \\
OGLE-SMC-02     &  0:40:28.10 & -73:44:46.5 & SMC128.5 & 15192  &  18.368$\pm$0.002 & 19.389$\pm$0.004 & 18.368$\pm$0.002$^\ast$ & 1.022$\pm$0.005$^\ast$\\
{\it (2005-SMC-001)} & & & & & & & & \\
& & & & & & & & \\
OGLE-SMC-03 & 1:02:37.50 & -73:22:25.9 & SMC111.7 & 18094 & 20.66$\pm$0.01 & 20.77$\pm$0.01 & 20.43$\pm$0.32 & 0.10$\pm$0.01\\
{\it (2008-SMC-001)} & & & & & & & & \\
& & & & & & & & \\
OGLE-SMC-04 & 1:00:59.32 & -73:14:17.1 & SMC106.3 & 33618 & 18.477$\pm$0.001 & 19.463$\pm$0.005 & 18.69$\pm$0.79 & 0.75$\pm$0.10$^\dagger$ \\
& & & & & & & & \\
\hline
\hline
\end{tabular}
\end{center}
$^\ast$ Source brightness and colour was set to this of the event's baseline, assuming no blending and dark lens (after \citealt{Dong2007})\\
$^\dagger$ Source colour was derived using EROS $B$-band data.
\end{table*}

\begin{table*}
\caption{Parameters of the standard Paczy{\'n}ski microlensing model fits to the OGLE-III SMC events. }
\label{tab:models}
\begin{tabular}[h]{lrrrrrr}
\hline
\multicolumn{7}{c}{OGLE-SMC-02}\\
parameter & \multicolumn{2}{c}{5-parameter fit} & \multicolumn{2}{c}{4-parameter fit} & \multicolumn{2}{c}{7-parameter fit}\\
\hline
$\t0$   \dotfill        & 3593.6  & $\pm0.02$  &  3593.7 & $\pm0.02$ & 3593.0 & $\pm0.02$ \\
& & & & & &\\
$\tE$   \dotfill        & 195.6  & $\pm1.9$ &  163.2 & $\pm0.3$ & 190.6 & $\pm1.6$ \\
& & & & & &\\
$\u0$  \dotfill        & 0.07132 & $\pm0.08208$ & 0.08915 & $\pm0.00016$ & 0.07340 & $\pm0.00071$ \\
& & & & & &\\
$\I0$  \dotfill     & 18.37 & $\pm0.002$  & 18.36 & $\pm0.002$ & 18.37 & $\pm0.002$ \\
& & & & & &\\
$\fsi$ \dotfill        & 0.8024 & $\pm0.0089$   &  1.0   & --- & 0.8248 & $\pm0.0078$ \\
& & & & & &\\
$\V0$  \dotfill        & ---   & ---          & ---    & --- & 19.39 & $\pm0.004$ \\
& & & & & &\\
$f_{\rm S_V}$ \dotfill & ---  & ---           &  ---   & --- & 0.9102 & $\pm0.0095$ \\
& & & & & &\\
$\chi^2$\dotfill  & 2836.7 &                     &  3259.0 &   & 3593.0 &  \\
& & & & & &\\
${\chi^2\over N_\mathrm{dof}}$\dotfill  & 2.32  &       &  2.66 &  & 2.62 &  \\
\hline
\multicolumn{7}{c}{OGLE-SMC-03}\\
parameter & \multicolumn{2}{c}{5-parameter fit} & \multicolumn{2}{c}{4-parameter fit} & \multicolumn{2}{c}{7-parameter fit}\\
\hline
$\t0$   \dotfill        & 4476.1  & $\pm0.1$  &  4476.1 & $\pm0.1$ & 4476.1 & $\pm0.1$ \\
& & & & & &\\
$\tE$   \dotfill        & 45.5  &$\pm$6.2 &  54.1 & $\pm0.9$ & 47.1 & $\pm4.4$ \\
& & & & & &\\
$\u0$  \dotfill        & 0.147 & $\pm$0.024 & 0.1161 & $\pm0.0013$ & 0.140 & $\pm0.016$ \\
& & & & & &\\
$\I0$  \dotfill     & 20.62 & $\pm0.01$  & 20.62 & $\pm0.01$ & 20.63 & $\pm0.01$ \\
& & & & & &\\
$\fsi$ \dotfill        & 1.32 & $\pm$0.26   &  1.0   & --- & 1.25 & $\pm$0.16 \\
& & & & & &\\
$\V0$  \dotfill        & ---   & ---          & ---    & --- & 20.77 & $\pm0.01$ \\
& & & & & &\\
$f_{\rm S_V}$ \dotfill & ---  & ---           &  ---   & --- & 1.20 & $\pm$0.16 \\
& & & & & &\\
$\chi^2$\dotfill  & 708.49 &                     &  710.65 &   & 784.13 &  \\
& & & & & &\\
${\chi^2\over N_\mathrm{dof}}$\dotfill  & 1.01  &       &  1.01 &  & 1.02 &  \\
\hline
\multicolumn{7}{c}{OGLE-SMC-04}\\
parameter & \multicolumn{2}{c}{5-parameter fit} & \multicolumn{2}{c}{4-parameter fit} & \multicolumn{2}{c}{7-parameter fit$^\ast$}\\
\hline
$\t0$   \dotfill        & 2611.6  & $\pm0.1$  &  2611.6 & $\pm0.1$ & 2611.6 & $\pm$0.1 \\
& & & & & &\\
$\tE$   \dotfill        & 18.60  & $^{+1.96}_{-1.85}$ &  17.17 & $\pm0.27$ & 18.3 & $\pm$1.8 \\
& & & & & &\\
$\u0$  \dotfill        & 0.3143 & $^{+0.0595}_{-0.0472}$ & 0.3589 & $\pm0.0039$ & 0.3231 & $\pm$0.050 \\
& & & & & &\\
$\I0$  \dotfill     & 18.477 & $\pm0.001$  & 18.477 & $\pm0.001$ & 18.477 & $\pm$0.001 \\
& & & & & &\\
$\fsi$ \dotfill        & 0.84 & $^{+0.22}_{-0.16}$   &  1.0   & --- & 0.87 & $^{+0.21}_{-0.16}$ \\
& & & & & &\\
$B_{\rm{0}}$  \dotfill        & ---   & ---          & ---    & --- & 19.48 & $\pm$0.01 \\
& & & & & &\\
$f_{\rm S_B}$ \dotfill & ---  & ---           &  ---   & --- & 1.03 & $^{+0.25}_{-0.18}$ \\
& & & & & &\\
$\chi^2$\dotfill      & 1052.3 &             &  1052.9 &   & 1154.6 &  \\
& & & & & &\\
${\chi^2\over N_\mathrm{dof}}$\dotfill  & 1.43  &       &  1.43 &  & 1.15 &  \\
\hline
\hline
\end{tabular}
\\
$^\ast$ fit to the OGLE $I$-band and EROS $B$-band data
\end{table*}

\subsection{OGLE-SMC-02}

This event was already known before this work as it was detected in real-time with the OGLE's Early Warning System (EWS) and was designated 2005-SMC-001\footnote{http://ogle.astrouw.edu.pl/ogle3/ews/2005/smc-001.html}.
It was probably one of the longest, the brightest and the best observed events found to date in the entire history of microlensing monitoring of both Magellanic Clouds.
Its light curve exhibited not only a clear parallax effect but also a deviation to the single microlensing model around its peak.
This anomaly is so tiny that it is not clearly seen in the OGLE data alone as shown in Fig.\ref{fig:events}.
However, its presence was explained with a non-caustic crossing binary lens model by \cite{Dong2007}.
In their analysis they also successfully employed for the first time space-parallax effect from additional observations taken by the Spitzer satellite, what helped break model degeneracies and solve the microlensing event.
\cite{Dong2007} concluded that the most likely location of the lens is the Galactic halo and with no light detected from the lens they suggested the lens is a binary black hole with a total mass of around 10$\msun$.
Nevertheless, they have not completely excluded the self-lensing scenario with both source and the lens residing in the SMC.

In our study we performed a simple modelling of a single point lens event with no parallax included. 
This was the model we used in the automated search pipeline, therefore we were able to derive the detection efficiency for this event. 
The values of the fitted parameters are gathered in Table \ref{tab:events} and the fit is shown in Fig. \ref{fig:events}.
The time-scale of $\te=190.6\pm1.6$ days we obtained for $I$ and $V$ data is conveniently in rough agreement with the time-scale derived by \cite{Dong2007} (between 160 and 190 days, depending on the details of the model configuration).

The blending parameters for both bands obtained in our model are relatively close to 1, however not including the parallax and the binarity in our model should affect this value severely. Full modelling of this event performed by \cite{Dong2007} (their Table 1) returned a blending fraction oscillating around zero-blending solutions, even suggesting some amount of negative blending in case of a few models. 

We analysed the astrometry of the residual images of this event obtained with the DIA and measured the centroid shift caused by blending during the event with respect to its position during the baseline as measured on the superb quality template image.
The accuracy of such a centroid shift can reach a few tens of milli-arc-seconds for bright events.
In case of OGLE-SMC-02 we did not detect any displacement of the centroid of light, which indicates there is no additional blending light  coming from a coincidental neighbouring star.
The lack of blending in microlensing models by \cite{Dong2007} was also confirmed by them with high resolution HST imaging and, along with other factors, led to the conclusion the lens in this event is either dark or very faint.

Cross-match of the lens position with available high energy data from ROSAT, XMM (F. Haberl \& R. Sturm, private communication) and INTEGRAL (A. Frankowski, private communication) returned no signal which can be associated with the black-hole. On the other hand, we would see the X-ray signal only if there was an accreting disk around the black-hole, the feature very unlikely to be present in a binary black-hole system.
Further detailed observations and studies are required in order to confirm the black hole origin of this event.

\subsection{OGLE-SMC-03}
This event was detected in real-time while it was on-going in 2008 by
the EWS (2008-SMC-001)\footnote{http://ogle.astrouw.edu.pl/ogle3/ews/2005/smc-001.html}.
It occurred on a faint blue star but was quite well covered in both
OGLE bands (see Fig. \ref{fig:events}), especially on the rising part
of the light curve which assures this was not a nova-like outburst.

In order to improve the quality of the light curve of this faint
object the original data were re-reduced so the flux variations were 
measured at the actual position of the event.
The data were then fit with a standard microlensing model and model
with parallax effect included as shown in Fig. \ref{fig:events}.
The parameters of the models are gathered in Tables \ref{tab:models}
and \ref{tab:parallax03} for standard and parallax models,
respectively.

All models had a strong indication of negative blending most likely
caused by an underestimation of the template flux of the faint source
and fringing present at the template image. This is not surprising for
the object fainter than 20th magnitude where unresolved background stars
can cause local overestimation or underestimation of the background level.
Removal of the lensed source following the method of
\cite{GouldAn2002} revealed no blended objects visible above the
background of the images.

The position of the source of the event shown on the CMD
(Fig.\ref{fig:cmd}) is based on the colour and brightness of the
source derived from the best microlensing fit with the negative
blending, but the uncertainty caused by this effect is encoded in the
large error-bar of the source brightness.

\begin{figure}
\center
\includegraphics[width=8.0cm]{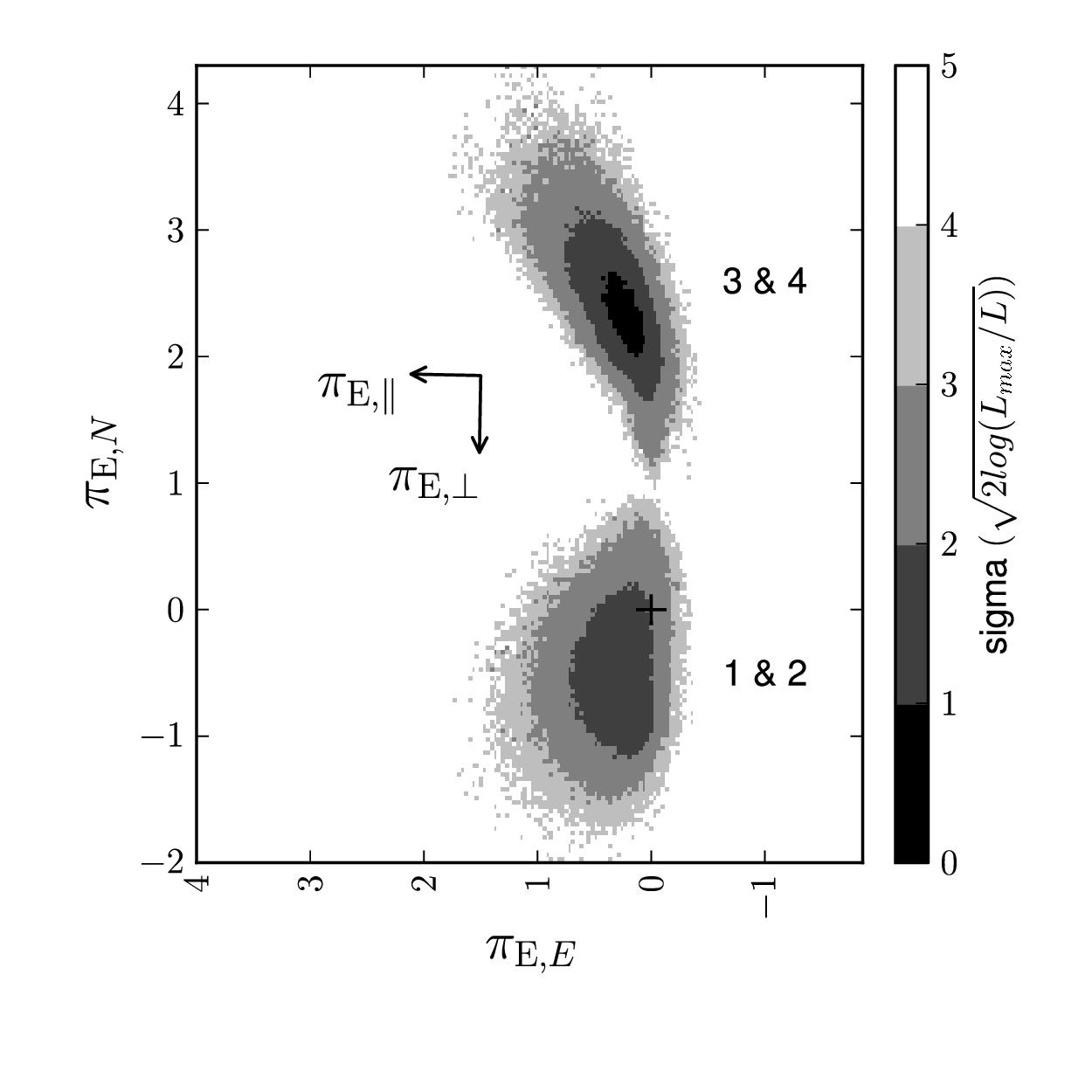}
\caption{Likelihoods of the North and East components 
of the parallax signal detected in OGLE-SMC-03 event.
Two regions corresponding to solutions 1 \& 2 and solutions 3 \& 4
are created through the jerk-parallax microlens degeneracy.
The directions parallel ($\parallel$) 
and perpendicular ($\perp$) to the Sun's apparent
acceleration in the geocentric frame are also shown.
(See Figure 3 in \citealt{Gould2004LMC}).
Grey-scale contour shows boundaries of $\sigma$ levels, which correspond
to the ratios of the likelihoods of the given region ($L$) to the most 
likely region ($L_{\rm max}$), as derived with MCMC.
Black cross mark origin of the coordinate system.}
\label{fig:parallax03}
\end{figure}

\begin{figure}
\center
\includegraphics[width=8.0cm]{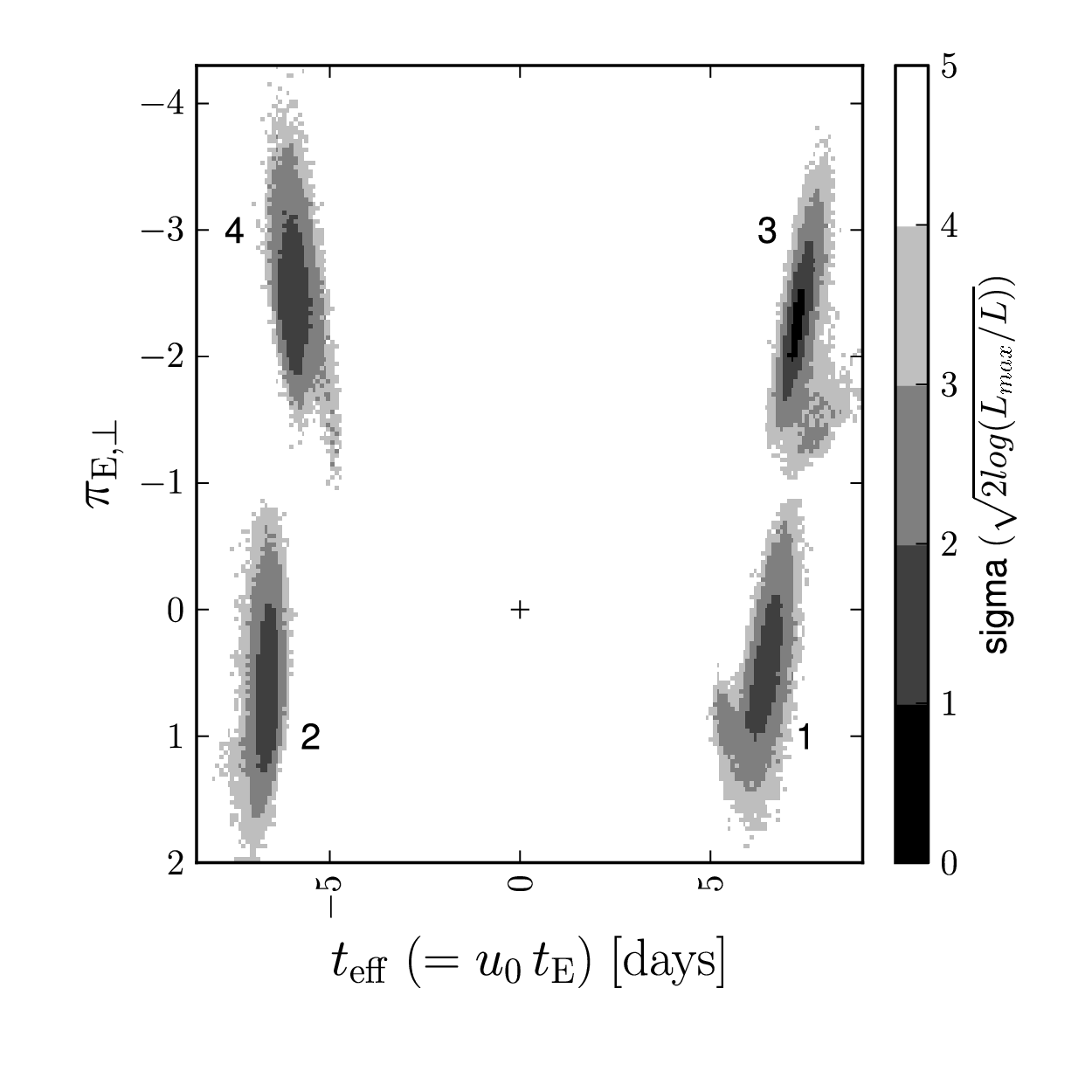}
\caption{Four microlensing solutions generated by
the jerk-paralax degeneracy and the $\u0$ degeneracy in 
the plane of $t_{\rm eff}$ and $\pi_{{\rm E}, \perp}$.
The  $\pi_{{\rm E}, \perp}$ is a projection of the 
parallax vector onto direction 
perpendicular to the Sun's acceleration in the geocentric frame
at the specified time $t_{0,par}$ -- this is the natural coordinate
for the jerk-parallax degeneracy. We use $t_{\rm eff}$ instead of 
$\u0$ since this parameter yields much smaller scatter.
Numbers indicate the individual solutions 
from Table \ref{tab:parallax03}.
}
\label{fig:foursolutions03}
\end{figure}

\begin{table}
\caption{Parameters of the microlensing models with parallax fit to
the event OGLE-SMC-03.}
\label{tab:parallax03}
\begin{center}
\begin{tabular}[h]{lcccc}
model no & 1 & 2 & 3 & 4\\
                     &    $\u0>0$ &    $\u0<0$ &    $\u0>0$ &    $\u0<0$ \\
& \multicolumn{2}{c}{($\pi_{\rm{E},\perp}>-\pi_{j,\perp}/2$)} & 
  \multicolumn{2}{c}{($\pi_{\rm{E},\perp}<-\pi_{j,\perp}/2$)} \\
\hline
$\t0$                &   4476.238 &   4476.242 &    4476.25 &   4476.241 \\
                     & $\pm$0.085 & $\pm$0.096 &  $\pm$0.11 & $\pm$0.072 \\
                     &            &            &            &            \\
$\tE$                &         47 &         54 &         56 &         41 \\
                     &    $\pm$21 &    $\pm$27 &    $\pm$28 &    $\pm$11 \\
                     &            &            &            &            \\
$\pi_{\mathrm{E},N}$ &      -0.43 &      -0.56 &       2.31 &       2.55 \\
                     &  $\pm$0.38 &  $\pm$0.40 &  $\pm$0.37 &  $\pm$0.39 \\
                     &            &            &            &            \\
$\pi_{\mathrm{E},E}$ &       0.33 &       0.37 &       0.35 &       0.30 \\
                     &  $\pm$0.23 &  $\pm$0.25 &  $\pm$0.25 &  $\pm$0.22 \\
                     &            &            &            &            \\
$\u0$                &      0.148 &     -0.140 &      0.145 &     -0.152 \\
                     & $\pm$0.036 & $\pm$0.039 & $\pm$0.039 & $\pm$0.033 \\
                     &            &            &            &            \\
$\I0$                &      20.62 &      20.62 &      20.62 &      20.62 \\
                     &  $\pm$0.01 &  $\pm$0.01 &  $\pm$0.01 &  $\pm$0.01 \\
                     &            &            &            &            \\
$\fsi$               &       1.41 &       1.35 &       1.49 &       1.47 \\
                     &  $\pm$0.34 &  $\pm$0.42 &  $\pm$0.36 &  $\pm$0.35 \\
                     &            &            &            &            \\
$\V0$                &     20.77 &     20.77 &     20.77 &     20.77 \\
                     & $\pm$0.01 & $\pm$0.01 & $\pm$0.01 & $\pm$0.01 \\
                     &            &            &            &            \\
$f_{\rm S_V}$        &       1.47 &       1.39 &       1.54 &       1.54 \\
                     &  $\pm$0.35 &  $\pm$0.42 &  $\pm$0.36 &  $\pm$0.38 \\
\hline                                         
$\chi^2$             &     780.80 &     780.80 &     780.49 &     780.39 \\
$\frac{\chi^2}{N_\mathrm{dof}}$ 
                     &     1.0180 &     1.0180 &     1.0176 &     1.0175 \\
\hline
\end{tabular}
\end{center}
\end{table}

Because the event was relatively long (it lasted for about 100 days)
we also investigated if the parallax model can reproduce the data
better than the standard model.
The modelling with MCMC found four equally valid parallax solution
covering all possible degeneracies.
Their $\chi^2$ for multi-band data fit, however, differed insignificantly ($\Delta \chi^2 \approx 0.4$, 
see Table \ref{tab:parallax03}) and were better by $\Delta \chi^2 \approx 3.4$ 
than the best non-parallax model with $\chi^2=784.13$.

We adopted the geocentric formalism to describe the parallax effects
(\citealt{An2002}, \citealt{Gould2004LMC}),
where we assume the reference system to be located at the Earth at the
time $t_{0,par}=2454476.23$, which we choose close to the peak of the
event.
This ensures that the values of $\u0$, $\t0$ and $\tE$ in
the parallax model are close to those from non-parallax model. 
Any deviations of the observer position due to Earth motion are calculated
against this reference system, which is in rectilinear motion at the
velocity of the Earth at that time.
The heliocentric velocity of the reference system projected on
the plane perpendicular to the line of sight is approximately
29 ${\rm km}\,{\rm s}^{-1}$ to South.
Usage of the geocentric system helps to find and identify 
all degenerate solutions.

To find all solutions we investigated the four-fold 
microlensing degeneracy that is created in case of a single 
lens with parallax from the two coexisting discrete degeneracies:
a two-fold $\u0$ degeneracy \citep{Smith2003parallax} 
and the jerk-parallax degeneracy \citep{Gould2004LMC}. 
These yield four solutions that are
presented in Figures \ref{fig:parallax03} and \ref{fig:foursolutions03}
 -- fit parameters for solutions 1 though 4 are 
gathered in Table \ref{tab:parallax03}.

Pairs of parallax solutions indicate either very close (solutions 3 \& 4)
or a bit more distant lens (solutions 1 \& 2). The former would yield 
a distance to the lens of order of 100 pc for a typical mass of the lens 
$\sim 0.3\, \msun$ -- it is extremely unlikely for the potential lens 
to be located so close to the observer. On the other hand, the latter set of 
solutions have parallax scale of $\pi_{\rm E}=0.69\pm0.32$.
This yields distance $2.2 \pm 4.3 {\rm kpc}$ for a typical mass, 
and since it covers significantly more volume in the Galaxy and 
many more potential lenses we choose these as preferred solutions.
The detection of the parallax signal in this event implies that 
the lens most probably comes from the Galactic thick disk population
of stars.

Assuming the microlensing source to be at the distance of the SMC, as
suggested by the CMD location of the source, we can neglect source
motion and calculate lens velocity projected on the
Earth plane to be ${\rm AU}\, \pi_{\rm E}^{-1}\, t_{\rm E}^{-1}$ $\approx$ 50
${\rm km}\,{\rm s}^{-1}$.
This, when taking into account the angle of the relative lens motion
and the velocity of the geocentric reference system, leads to heliocentric lens velocity,
projected on the plane of the observer, to be
approximately $(-70, 30)$ ${\rm km}\,{\rm s}^{-1}$ in North and East
direction, respectively.
Since the lens is much closer than the source we can neglect the
effect of projection and note that the true heliocentric velocity of
the lens will be of the same order.
This type of kinematics is not surprising for the thick disk object.

Because there is no blend visible we can derive an upper limit on the
lens brightness equal to around 21.5 mag in $I$-band.
This limit is met by a late M dwarf (with mass $\lesssim 0.2 \msun$)
at the distance of more than $1.5$ kpc, and an early M dwarf (with
mass $\lesssim 0.3 \msun$) at the distance of more than $5$ kpc.
Although we cannot rule out the lens to be a dark object in the 
Galactic halo the lens as a disk star is a much more natural explanation.
\cite{CalchiNovati2011} showed in their simulation of the
microlensing rate towards the LMC that the contribution of the Galactic 
Disc lenses to the overall microlensing rate is comparable to 
the contribution from the LMC self-lensing. It means that in our 
searches for microlensing events towards Magellanic Clouds 
we should expect some of them to be caused by Galactic lenses.

We can conclude that the most likely scenario for OGLE-SMC-03 is that the
lens is an M dwarf with mass of about 0.1-0.3 $\msun$ located 1-5 kpc
from the Earth and belonging to the thick disk of the Galaxy, similar to the event MACHO-LMC-5, e.g. \cite{Drake2004}, \cite{Gould2004LMC} 
and event EROS2-LMC-8 \citep{TisserandEROSLMC}.

With velocity of the order of 80 ${\rm km}\,{\rm s}^{-1}$ and location
a few kpc from the Earth, the lens should have a noticeable proper
motion of a few ${\rm mas}\, {\rm yr}^{-1}$ and therefore should be
resolvable with the {\it HST} in a couple of years.
This would give a good opportunity to confirm the nature of this event.

\subsection{OGLE-SMC-04}
This event was not previously detected neither by EWS (it occurred in the end of 2002 in the early season of the OGLE-III when the EWS was not yet operating), nor by any other surveys.
Its light curve is relatively well covered by OGLE-III $I$-band data, but there are no observations taken during the event in $V$-band.

The event was present in the EROS2 database, however only in their $B$-band. 
Therefore we performed a multi-band fit using OGLE $I$-band and EROS $B$-band data, which resulted in instrumental colour of the source of $(B-I)_{\rm{S}}=0.82\pm0.04$. This was then transformed to standard OGLE's $V-I$ following \cite{TisserandEROSLMC} and yielded $(V-I)_{\rm{S}}=1.02\pm0.08$. The error-bar includes the systematic error from the transformation.
The position of the source on the CMD (Fig.\ref{fig:cmd}) indicates it belongs to the Red Clump giant population of the SMC.


The time-scale of the event is relatively short ($\tE=18.3\pm1.8$), which favours a self-lensing (SL) scenario. 
Unless the lens is nearby ($D_L<$20 kpc), a typical time-scale of halo lensing events with masses around 0.4 $\msun$ and sources in the SMC are well above 20 days.
In SL the time-scales are usually a few times shorter owing to much larger projected velocity, $\tilde{v}_{SL}\approx$ 2000 km/s,
compared to $\tilde{v}_{halo}\approx$ 200 km/s for halo lenses \citep{BoutreuxGould1996}.
This places this event as a good candidate for a self-lensing microlensing interpretation.

Another possibility if that the lens is a thick disk red dwarf given the fact the blending object must be redder that the Red Clump of the SMC. High-resolution imaging could potentially reveal the actual lens as the event happened already almost 10 years ago giving enough time for lens and source to separate. 

\section{Blending and detection efficiency}
\label{sec:blending}

\begin{figure}
\includegraphics[width=8.5cm]{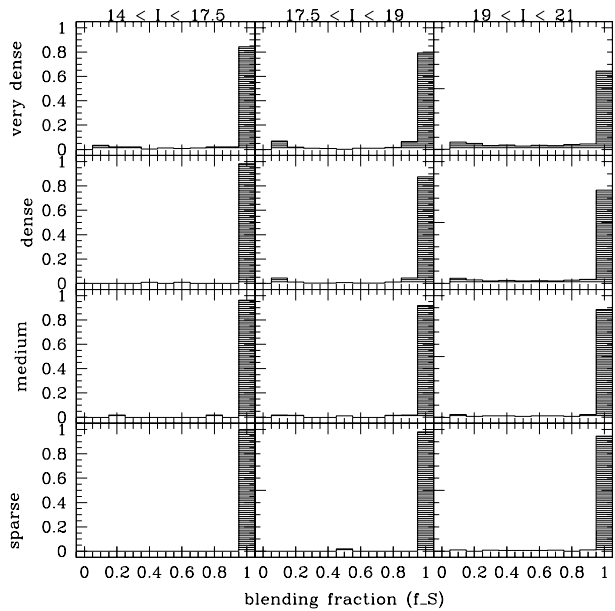}
\caption{
Distributions of blending parameter (source flux fraction in baseline flux) for selected levels of stellar density derived for the simulated OGLE--III SMC images based on deep archival HST images. 
The distributions are shown for three magnitude bins.
The density levels shown correspond to stellar densities of $\log(N_*/\rm{CCD~chip})=(4.9,4.6, 4.4,3.9)$ for very dense, dense, medium and sparse fields, respectively.
}
\label{fig:blending}
\end{figure}

\begin{figure}
\includegraphics[width=8.5cm]{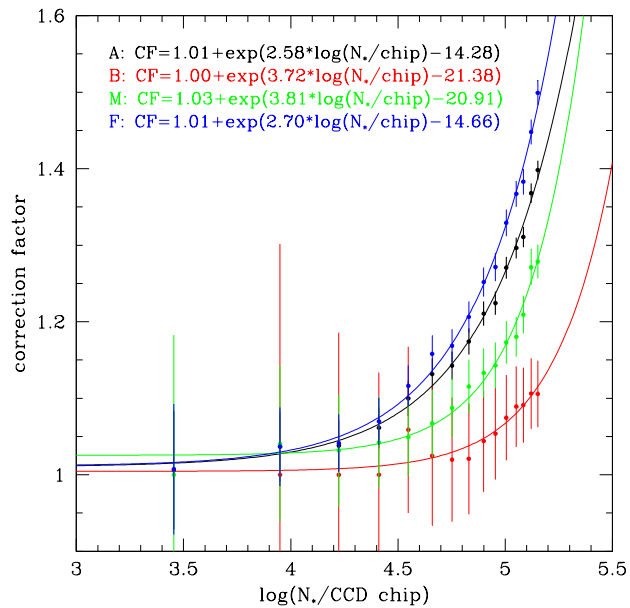}
\caption{
Correction factor (CF) for the number of monitored stars as a function of the density of the stars on a single CCD chip of the OGLE-III template image. Shown are the CFs for three magnitude bins (B for bright, M for medium and F for faint) and for the entire magnitude range (A for all). The legend shows the expression for the curve fit to each of the data sets.
}
\label{fig:cf}
\end{figure}

\begin{figure}
\includegraphics[width=8.5cm]{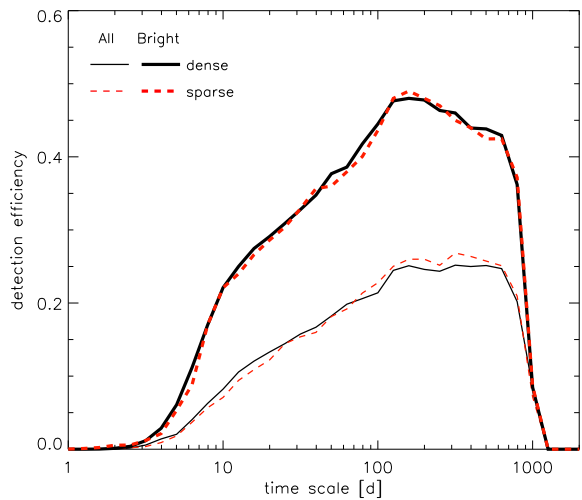}
\caption{
Microlensing events detection efficiency for example dense (solid lines, field SMC106.3) and sparse (dashed lines, field SMC111.7) fields of the SMC as observed by OGLE-III. Shown are curves for All Stars Sample and Bright Stars Sample (thin and thick lines, respectively). 
}
\label{fig:eff}
\end{figure}

As in the previous parts of the study of the OGLE microlensing data towards the Magellanic Clouds we carefully tackled the issue of blending in these crowded fields.
Here we applied the method developed for Paper III.
In brief, we simulated OGLE-III images for a range of stellar densities using deep luminosity functions of three representative fields from the {\it HST} Local Group Stellar Photometry archive \citep{Holtzman2006}. Locations of the selected fields are marked on Fig. \ref{fig:fields}.
Then we compared simulated OGLE images with the input catalogue of stars and were able to assign a number of real {\it HST} stars to each blended object detected on a simulated frame.
For each density level and in three magnitude bins, $14<I<17.5$ (bright), $17.5<I<19$ (medium) and $19<I<21$ (faint) we derived a distribution of flux ratio between each {\it HST} component and the blended object it was residing in. 
The blending distributions are shown in Fig. \ref{fig:blending}.
We were also able to calculate a mean ratio between the number of real stars to all objects composed of these stars. 
This ratio was then used as a correction factor (CF) for observed OGLE template images in order to estimate the real number of monitored stars in each field (see Table \ref{tab:fields}).
The CF is shown in Fig. \ref{fig:cf} for all stars (A) and three magnitude bins: bright (A), medium (M) and faint (F).

With these blending distributions at hand we were able to perform the detection efficiency determination.
For selected fields representing different density levels and for time-scales in range between 1 and 1000 days we simulated numerous microlensing events with the parameters being drawn from realistic distributions. 
For the lensed flux we used the measurements of randomly selected star from the simulated field, which during the event was split between blend and source according to the blending parameter. This procedure also preserved any variability of the underlying flux, allowing for events with variable baseline (see \eg \citealt{varbaseline}). 

The efficiency curves for representative dense and sparse fields and for the All Stars and Bright Stars samples are shown in Fig. \ref{fig:eff}. 
The detection efficiencies do not change much with the density of the field due to relatively low gradient in the stellar density in the OGLE-III SMC fields. 
The calculated efficiency is somewhat lower when compared with the one obtained for OGLE-III LMC data, mainly because here we used a different magnitude cut on star selection for the All Stars Sample (21.0 mag, compared with 20.4 mag for the LMC). 


\section{Optical depth}
Having the three events and their time-scales as well as the detection efficiency of each of them we are able to calculate the average optical depth towards the SMC based on OGLE-III data.
In order to evaluate Eq. \ref{eq:tau} we also need the estimated number of total monitored stars, derived in the previous section, and the total time of observation, $T_{\rm obs}$=2870 days. 

First, we calculated the optical depth for the All Stars Sample ($N_*=5\;971\;776$, down to $I$=21 mag) using the detection efficiencies obtained in our simulations. 
Because our simulations utilised $I$-band data solely we used $\te$ obtained in 5-parameter fit for all three events.
This led to total $\tau_{\rm SMC-OIII}=1.17\pm 0.91~\times~10^{-7}$. The error on $\tau$ was calculated following \cite{HanGouldtau} and reflects only uncertainty caused by the small number of events used.

Then, similarly as in Papers I, II and III, we corrected the efficiency for the fact our simulations were only considering single-point events and that lensing by a binary (and other exotic effects) was not included.
Efficiencies were scaled down by 10 per cent and the optical depth for efficiency corrected for binary events was $\tau_{\rm SMC-OIII}=1.30\pm1.01~\times~10^{-7}$. Details of the calculations for each event are shown in Table \ref{tab:tau}.

For the Bright Stars Sample ($N_*=1\;702\;724$, down to $I$=19.3 mag) the optical depth calculated for two events (\#2 and \#4) is $\tau_{\rm SMC-OIII}=1.52\pm 1.40~\times~10^{-7}$ and $\tau_{\rm SMC-OIII}=1.69\pm 1.55~\times~10^{-7}$, when efficiencies were corrected for lack of sensitivity for binary events. Within the (large) error-bars, the estimates of the optical depth for the All Stars and Bright Stars Samples are in agreement. 
However, it is clearly seen that higher detection efficiency does not completely counterbalance the smaller number of stars and smaller number of events in the Bright Stars Sample.

\begin{table}
\caption{The optical depth for the three events found in the OGLE--III SMC data calculated for the All Stars Sample. The columns show event name, its time-scale, detection efficiency and individual event's contribution to the total optical depth.}
\label{tab:tau}
\begin{center}
\begin{tabular}[h]{cccc}
\hline
event & $\te$ & $\epsilon(\te)$ &$\tau_i \times 10^{-7}$ \\
\hline
\multicolumn{4}{c}{efficiency not corrected for binary events} \\
& & & \\
OGLE-SMC-02$^\dagger$ & 195.6$\pm$1.9 & 0.234942 & 0.76\\
& & & \\
OGLE-SMC-03$^\dagger$ & 45.5$\pm$6.2  & 0.155922 & 0.27 \\
& & & \\
OGLE-SMC-04 & 18.60$^{+1.96}_{-1.85}$ & 0.123365 & 0.14 \\
\hline
total $\tau_{\rm SMC-OIII}$   &              &          &   $1.17\pm0.91$ \\
\hline
\hline
\multicolumn{4}{c}{efficiency corrected for binary events} \\
& & & \\
OGLE-SMC-02$^\dagger$ & 195.6$\pm$1.9 & 0.211448  &  0.85 \\
& & & \\
OGLE-SMC-03$^\dagger$ & 45.5$\pm$6.2 & 0.140330  & 0.30 \\
& & & \\
OGLE-SMC-04 & 18.60$^{+1.96}_{-1.85}$ & 0.111028  &  0.15 \\
& & & \\
\hline
total $\tau_{\rm SMC-OIII}$   &              &          &   $1.30\pm1.01$ \\
\hline
\end{tabular}
\\
\end{center}
$^\dagger$ Time-scale and efficiency for these events are taken from a standard fit.
\end{table}

Note that the binary-lens event OGLE-SMC-02 was included in the optical depth calculations above as if it were a standard event.
This approximation is justified because even though the light curve of this event deviates a little from the standard ``Paczy{\'n}ski'' curve, the event passes through our automated pipeline, which does not consider any exotic effects. 
Moreover, the time-scale obtained in the standard fit is very close to the one obtained in the full model by \cite{Dong2007}.
Therefore it was also possible to derive the detection efficiency for this event.
The same applies to the parallax event OGLE-SMC-03

\section{Discussion}

In the 8 years of OGLE-III data covering the Small Magellanic Cloud and its surroundings we detected three convincing microlensing events candidates.

This is definitively the best set of candidates presented in the series of our papers concerning microlensing searches towards the Magellanic Clouds with OGLE data. 
The microlensing nature of all three events is very difficult to disprove - none of their sources is located in the region of potential contaminants ``Blue Bumpers'' (Fig. \ref{fig:cmd}) and their light curves covering at least 8 years (12 years for OGLE-SMC-02 as it was also monitored during the OGLE-II) show no other additional bumps. Additional data spanning for a couple of more years available for all three events from the MACHO and EROS groups also show no further bumps.

However, still the most difficult question to answer is where these events originated from.
In principle, because standard microlensing model can not tell us where the source and the lens are located, there are a number of different combinations possible, including the SMC, the halo of the Milky Way and the disk of the Milky Way.
In the history of the Magellanic Clouds microlensing there were already examples of confirmed cases of each of the combinations, \eg source and lens from the SMC \citep{Assef2006MACHO97SMC1}, disk lens towards the LMC \citep{Kallivayalil2004}. 

In the case of our events, the new one, OGLE-SMC-04, is probably the best candidate for self-lensing event given its time-scale and location in the Red Clump stars of the SMC, however, we can not exclude the thick disk red dwarf scenario or MACHO. 
The parallax effect detected in the event OGLE-SMC-03 help constrain the lens to most likely be a typical M-dwarf star from the Galactic disk, however a dark lens can not be entirely ruled out.
Finally, even though OGLE-SMC-02 is the likely to be a halo lens, \cite{Dong2007}, who modelled it in detail, did not exclude the self-lensing option completely.

The effect of self-lensing for the SMC is not yet well understood.
However, due to the alignment of the SMC along the line-of-sight the contribution of SL to $\tau$ is expected to be higher than in the face-on LMC and be at least 1.0$\times 10^{-7}$, in average (\citealt{EROSSMC1998}, \citealt{Graff1999}).
This is in good agreement with the value of the optical depth we measured for all three events found in the OGLE-III data.
Therefore, we expect the signal from the self-lensing or, more generally, the background of lensing by known stellar populations, is close to what we have detected. 
However, because the nature of our events is not yet firmly confirmed we can derive an upper limit on dark matter compact halo objects, following \cite{AlcockMACHOBH2001}, \cite{TisserandEROSLMC} and Papers I, II and III.

\begin{figure*}
\includegraphics[width=13.5cm]{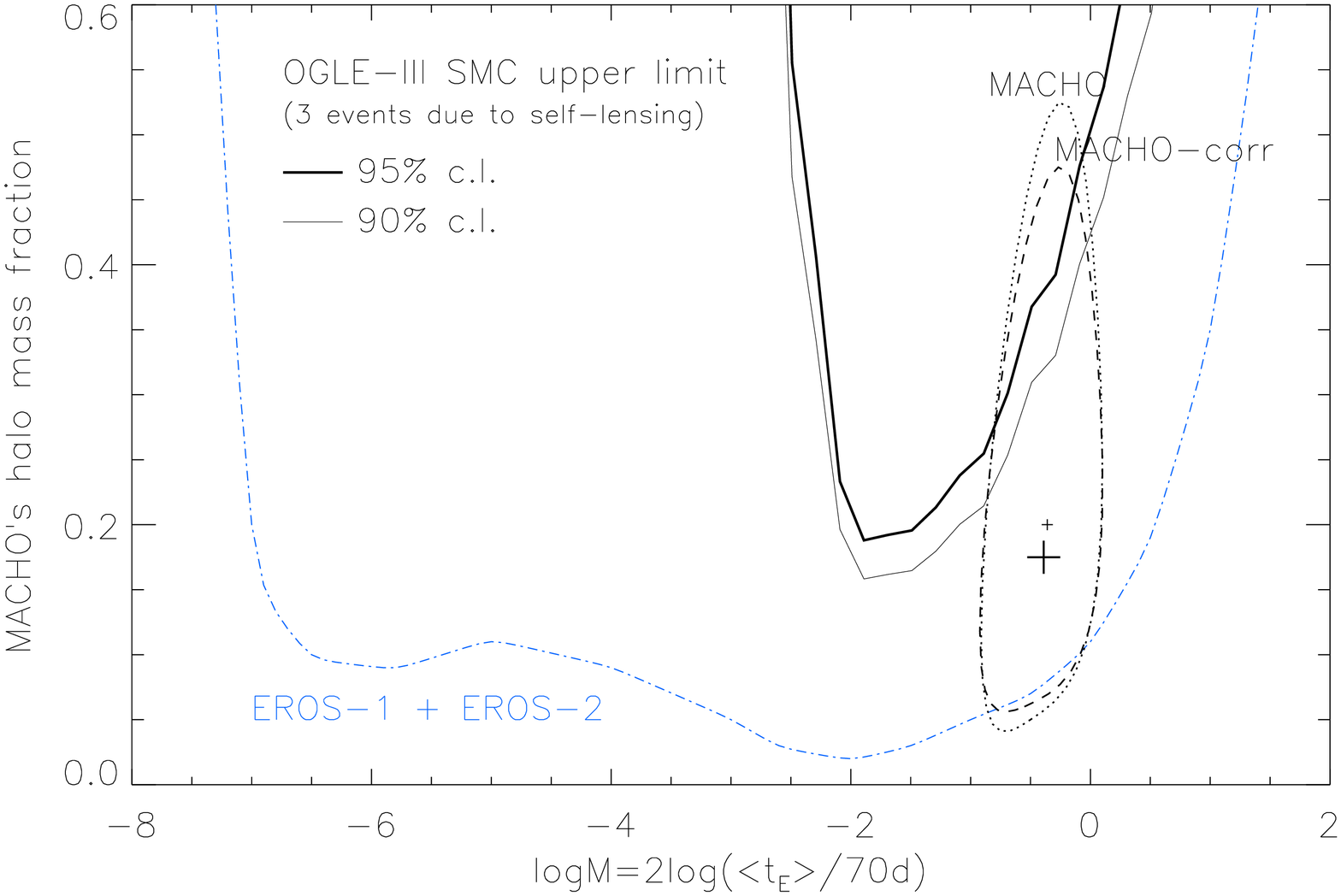}
\caption{Fraction of the mass of MACHOs in the halo as derived from OGLE-III SMC data. Black solid curves show an upper limit from the OGLE-III SMC data assuming the background self-lensing signal of three events. Also shown is the upper limit from EROS (blue dashed curve), original measurement by MACHO (small cross and dotted curve) and the MACHO result corrected for the fact that one of their events was rejected by OGLE-III data (dashed curve, big cross). 
}
\label{fig:taulimitSMC}
\end{figure*}

Assuming the MACHOs mass distribution function from model ``S'' from \cite{AlcockMACHOLMC} and using the mean detection efficiency for our SMC fields we derived the number of expected events due to MACHOs as a function of their mass. 
The mass in such events can be approximately translated to a time-scale as $\log M=2\log (\tE/70)$. 
The number of expected events compared with the observed (or not observed) signal can be converted to an upper limit on the MACHO's halo mass fraction using enhanced Poisson statistics with background signal \citep{FeldmanLowStatistics}.
The derived upper limit for OGLE-III SMC data is shown in Fig. \ref{fig:taulimitSMC} and was calculated assuming we expect about 3 events from  the self-lensing/background.
For a ``typical'' MACHO with mass of 0.4 $\msun$ the fraction is less than 37 per cent, but for lower masses, between 0.01 and 0.1 $\msun$ it is below 20 per cent.
Figure \ref{fig:taulimitSMC} shows also an upper limit on MACHOs derived by EROS \citep{TisserandEROSLMC} and MACHO signal claimed by the MACHO group at 95 per cent confidence.
For the latter we show both the original value from \cite{BennettMACHOLMC} (for $\tau=1.0\pm0.3\times 10^{-7}$) for all MACHO events being due to MACHOs and the value corrected following the rejection of one of the original MACHO events (MACHO-LMC-7, see Paper III).
For the corrected value, the MACHO fraction in the halo becomes 18 per cent, compared to the original 20 per cent.
If the background lensing (Galactic disk and LMC self-lensing) was extracted from the MACHO signal (at a conservative level suggested in \citealt{BennettMACHOLMC} of $0.24\times10^{-7}$) and the compromised event was rejected, the MACHO group data can be interpreted that only about 14 per cent of MACHOs may compose the Galactic halo.

\begin{figure*}
\includegraphics[width=13.5cm]{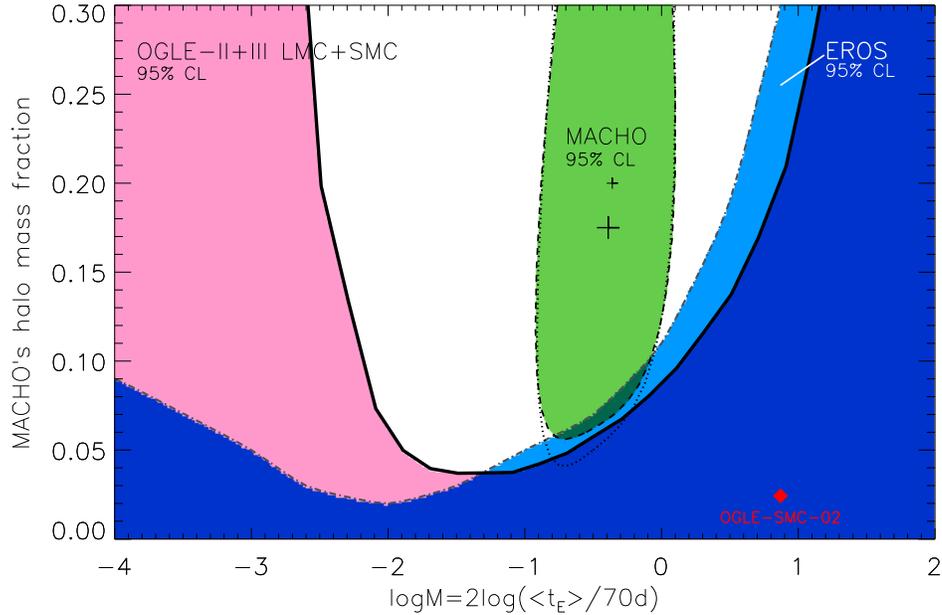}
\caption{Inclusion region for the fraction of the mass of MACHOs in the halo for combined OGLE-II and OGLE-III data for LMC and SMC (solid curve, pink/dark blue) when all OGLE events are attributed to expected self-lensing/background signal. The EROS upper limit and the MACHO signal are shown as in Fig.\ref{fig:taulimitSMC} in light/dark blue and green, respectively. Also shown is the fraction of mass due to BH-candidate event OGLE-SMC-02.
}
\label{fig:taulimit}
\end{figure*}

We can obtain a much tighter limit from OGLE by combining all OGLE-II and OGLE-III results for both Large and Small Magellanic Clouds. 
For each set of data we computed the expected number of events for a given mass of the deflector and combined it with about 8 events expected to be due to the background lensing populations.
It yielded $f<0.06$ (at 95 per cent C.L.) for masses around 0.4 $\msun$ and reached the minimum of $f<0.04$ at mass range between 0.01 and 0.15 $\msun$. 
Figure \ref{fig:taulimit} shows the combined OGLE limit along with the corrected MACHO group signal and EROS upper limit with the both axes zoomed in at the intersection of results from all surveys.
This result, therefore, pushes the upper limit on MACHO mass fraction in the halo down to a value similar to the one derived by EROS.
The OGLE result is a somewhat more sensitive to MACHOs in the higher mass regime (between 1 and 10 $\msun$), and is significantly less sensitive for masses below 0.01 $\msun$. 

When looking at all events in more detail, in the most likely case, we can safely attribute to non-MACHO lensing (self-lensing or disk lensing) all the events detected in the LMC and SMC during OGLE-II and OGLE-III except OGLE-SMC-02. 
In such case, the optical depth due to this single dark matter event would be $\tau_{SMC\#2}\approx 0.12\times 10^{-7}$.
If the MW halo were composed only from such lenses with mass of around 10 $\msun$, such an optical depth would mean they contribute no more than 2 per cent of the total mass of the dark halo.
This is in agreement with the expected mass fraction of black-holes in the total mass budget of the Galaxy varying from around 5 per cent \citep{Sartore2010} for all massive stellar remnants to 0.4 per cent for a standard initial mass function (\eg \citealt{Bastian2010}) integrated above 6 $\msun$.
It is also a much tighter limit than the limit of 40 per cent derived by the MACHO group on objects below 30 $\msun$ \citep{AlcockMACHOBH2001}, obtained based on a lack of long-term events. 
The value obtained for a single microlensing event, however, is only crudely estimated, and should be assessed more elaborately, \eg with simulations of long events in the OGLE-II and OGLE-III data combined. 
In practice, more events with long time-scales are needed in order to put better and more statistically sound constraints on the massive stellar remnants abundance in the halo. 
Continued observations of the Magellanic Clouds by MOA and OGLE-IV surveys will hopefully yield more potential black-hole events residing in the halo if they are common enough. 

\section{Conclusions}
We analysed OGLE-III data covering the Small Magellanic Cloud and found three convincing candidates for microlensing events, among which one was not known before (OGLE-SMC-04).
Two of the events are likely to be caused by self-lensing or have lenses located in the Galactic disk. 
OGLE-SMC-03 is actually the first event towards the Small Magellanic Cloud exhibiting clear parallax signal, which makes the lens likely to be in the thick disk. 
The third one (OGLE-SMC-02) is a potential candidate for a binary black hole lens from the halo, but its self-lensing origin can not be totally excluded.

Following a detailed analysis of the blending in the SMC we derived the optical depth for the All Stars Sample of $\tau_\mathrm{SMC-OIII}=1.30\pm1.01\times 10^{-7}$, which is in agreement with expected self-lensing signal towards the SMC.
The upper limit on the dark matter in form of MACHOs derived for the OGLE-III SMC data alone was about 20 per cent for masses below 0.1 $\msun$. 

In this paper we concluded the studies of the microlensing searches conducted so far by the OGLE project towards the Magellanic Clouds. 
We presented the final result combining all available data and deriving new constraints on the fraction of massive compact halo objects in the Galactic halo of 6 per cent for $M=0.1-0.4~\msun$ and below 4 per cent for lower masses. 
For MACHOs with mass of 1 $\msun$ and 20 $\msun$ the upper limits are $f<9$ and $f<20$ per cent, respectively. 
13 years of OGLE observations indicate the MACHO halo fraction is well below the value suggested by MACHO group (14 per cent  after correction due to rejection of one of their events owing to a second bump for the same events found in the OGLE data) and in agreement with the limit derived by EROS survey.

Our result indicates that baryonic dark matter in the form of relics of stars and very faint objects in the sub-solar mass regime is unlikely to inhabit the Milky Way's dark matter halo in any significant numbers.
The presence of the black hole lens candidate towards the SMC agrees with the expected $< 2$ per cent contribution of black holes to the mass of the Galactic halo.

With the OGLE project now in its fourth phase we hope the sensitivity to extremely long events will improve significantly within the next years when combined with the historic OGLE data.
Long duration events will also be easily detectable in near-real-time by Gaia satellite, due to launch in 2013. 
Detailed follow-up observations of the alerts triggered by OGLE and Gaia will be crucial for revealing the true nature of detected events.
This should result in an increase in the number of potential black-hole lenses or allow us to rule out heavy dark matter compact objects as well and close that topic definitively.


\section*{acknowledgements}
We would like to thank the referee, Andy Gould, for comments which improved the paper significantly. 
We also would like to thank for their help at various stages of this work Drs Nicholas Rattenbury, Vasily Belokurov, Martin Smith and Subo Dong. We also thank Drs. Adam Frankowski, Frank Haberl and Richard Sturm for their help with high-energy data.
This work was partially supported by EC FR7 grant PERG04-GA-2008-234784 to {\L}W.
JS acknowledges support through the Polish MNiSW grant no. N20300832/070 and Space Exploration Research Fund of The Ohio State University.
The OGLE project acknowledges funding received from the European Research Council under the European Community's Seventh Framework Programme (FP7/2007-2013), ERC grant agreement no. 246678.

\bibliographystyle{mn2e}

\label{lastpage}

\end{document}